\documentclass[epj]{svjour}
\usepackage{psfrag}

\usepackage{epsfig}

\def\lsim{\mathrel{\rlap{\lower4pt\hbox{\hskip1pt$\sim$}}
    \raise1pt\hbox{$<$}}}                
\def\gsim{\mathrel{\rlap{\lower4pt\hbox{\hskip1pt$\sim$}}
    \raise1pt\hbox{$>$}}}                
\begin{document}

%
\title{{\boldmath$DN$} interaction from meson exchange} 

\author{J. Haidenbauer\inst{1,2}, G. Krein\inst{3}, Ulf-G. Mei{\ss}ner\inst{1,2,4},
 and L. Tolos \inst{5,6}}

\institute{
Institut f\"ur Kernphysik and J\"ulich Center for Hadron Physics,
Forschungszentrum J\"ulich, D-52425 J\"ulich, Germany
\and Institute for Advanced Simulation,
Forschungszentrum J\"ulich, D-52425 J\"ulich, Germany
\and
Instituto de F\'{\i}sica Te\'{o}rica, Universidade Estadual
Paulista,
Rua Dr. Bento Teobaldo Ferraz, 271 -  01140-070 S\~{a}o Paulo, SP, Brazil 
\and
Helmholtz-Institut f\"ur Strahlen- und Kernphysik (Theorie)
and Bethe Center for Theoretical Physics,
Universit\"at Bonn, Nu\ss allee 14-16, D-53115 Bonn, Germany 
\and
Theory Group, KVI, Zernikelaan 25, NL-9747AA Groningen, The Netherlands
\and
Instituto de Ciencias del Espacio (IEEC/CSIC), Campus Universitat 
Autonoma de Barcelona, Facultat de Ciencies, Torre C5, E-08193 Bellaterra 
(Barcelona), Spain
}
\date{Received: date / Revised version: date}

\abstract{
A model of the $D N$ interaction is presented which 
is developed in close analogy to the meson-exchange $\bar KN$ 
potential of the J\"ulich group utilizing SU(4) symmetry constraints. 
The main ingredients of the interaction are provided by vector meson 
($\rho$, $\omega$) exchange and higher-order box diagrams involving 
${D}^*N$, $D \Delta$, and ${D}^*\Delta$ intermediate states. The
coupling of $DN$ to the $\pi\Lambda_c$ and $\pi\Sigma_c$ channels is taken
into account. 
The interaction model generates the $\Lambda_c$(2595) resonance 
dynamically as a $DN$ quasi-bound state. 
Results for $DN$ total and differential cross sections are presented
and compared with predictions of an interaction model that
is based on the leading-order Weinberg-Tomozawa term.
Some features of the $\Lambda_c$(2595) resonance are discussed and 
the role of the near-by $\pi\Sigma_c$ threshold is emphasized. 
Selected predictions of the orginal $\bar KN$ model are reported too.
Specifically, it is pointed out that the model generates two poles
in the partial wave corresponding to the $\Lambda$(1405) resonance.
}


\authorrunning{J. Haidenbauer et al.}
\titlerunning{$DN$ interaction from meson-exchange}

\maketitle

\section{Introduction}
\label{sec:intro}

The study of the low-energy interactions of open charm $D$-mesons with nucleons 
is challenging for several reasons. The complete lack of experimental data on 
the interaction makes it very difficult to constrain models. Reliable models
are very important for guiding planned experiments by the 
$\bar{\rm P}$ANDA~\cite{Lehmann:2009dx} and CBM~\cite{Staszel:2010zz} 
collaborations of the future FAIR facility at Darmstadt~\cite{FAIR}. 
Estimates for the magnitudes of cross sections are required for the design of 
detectors and of efficient data acquisition systems. As emphasized in recent
publications~\cite{Haiden07,Haiden08}, one way to make progress in such a
situation is to build models guided by analogies with other similar 
processes, by the use of symmetry arguments and of different dynamical 
degrees of freedom. 

Physics motivations for studying the interaction of $D$-mesons with nucleons
are abundant. Amongst the most exciting ones is the possibility of studying 
chiral symmetry in matter. The chiral properties of the light quarks composing 
$D$-mesons are sensitive to temperature and density; one expects to learn 
about manifestations of such a sensitivity through the detection of changes in 
the interaction of these mesons with nucleons in the medium as compared to the 
corresponding interaction in free space. Also, studies of $J/\psi$ dissociation 
in matter, since long time~\cite{Matsui:1986dk} suggested as a possible signature 
for the formation of a quark-gluon plasma, require a good knowledge of the 
interaction of $D$-mesons with ordinary hadrons~\cite{Thews} in order to 
differentiate different scenarios and models in this area. Another exciting 
perspective is the possibility of the formation of $D$-mesic 
nuclei~\cite{Tsushima:1998ru,GarciaRecio:2010vt} 
and of exotic nuclear bound states like $J/\psi$ binding to 
nuclei~\cite{Brodsky:1989jd,Luke92}. 
In this latter case in particular, the interaction of $D$-mesons with ordinary matter 
plays a fundamental role in the properties of such exotic
states~\cite{Krein:2010vp,Ko:2000jx}. 
  
In a recent paper we examined the possibility to extract information
about the $DN$ and $\bar DN$ interactions from the $\bar p d \to D^0D^-p$
reaction \cite{Haiden08}. 
The scattering amplitudes for $DN$ and $\bar DN$ employed in this exploratory 
study were generated from interaction models that were constructed along
the line of the $\bar KN$ and $KN$ meson-exchange potentials developed by 
the J\"ulich group some time ago \cite{MG,Juel1,Juel2}.
While the $\bar DN$ interaction has been described in some 
detail in Ref.~\cite{Haiden07} this has not been done so far for
the $DN$ model. With the present paper we want to remedy this situation.

The $DN$ interaction is derived in close analogy to the meson-exchange 
$\bar K N$ model of the J\"ulich group \cite{MG}, 
utilizing as a working hypothesis SU(4) symmetry constraints, 
and by exploiting also the close connection between 
the $DN$ and $\bar DN$ systems due to G-parity conservation. 
In particular, the $D N$ potential is obtained by substituting the
one-boson-exchange ($\sigma$, $\rho$, $\omega$) contributions, 
but also the box diagrams involving
$\bar K^*N$, $\bar K\Delta$, and $\bar K^*\Delta$ intermediate states,
of the original $\bar KN$ model of the J\"ulich group by the corresponding 
contributions to the $D N$ interaction.
Of course, we know that SU(4) symmetry is strongly broken, as is reflected in
the mass differences between the strangeness and the charm sectors, for 
mesons as well as for baryons. However, as already argued in 
Ref.~\cite{Haiden07}, we expect that the dynamics in the $DN$ ($\bar DN$) and
$\bar KN$ ($KN$) systems should be governed predominantly by the same 
``long-range'' physics, i.e. by the exchange of ordinary (vector and 
possibly scalar) mesons. Thus, in both systems the dynamics should involve primarily
the up- and down quarks, whereas the heavier quarks (the $s$ and $c$ quarks,
respectively) behave more or less like spectators and contribute predominantly 
to the static properties of the mesons and baryons. Therefore, the assumption of
SU(4) symmetry for the dynamics seems to be not completely implausible. 

In any case, invoking SU(4) symmetry for the $DN$ interaction involves 
larger uncertainties than for $\bar DN$. For the former, like in the 
analogous $\bar KN$ system, there are couplings to several other
channels which are already open near the $DN$ threshold ($\pi \Lambda_c$, $\pi\Sigma_c$)
or open not far from the threshold ($\eta \Lambda_c$).
The coupling to those channels should play an important role 
for the dynamics of the $DN$ system -- as it is the case in the 
corresponding $\bar KN$ system -- and, thus, will have an impact on the 
results, at least on the quantitative level. Specifically, the transitions 
from $DN$ to those channels involve the exchange of charmed mesons, for example 
the $D^*$(2010), where the range arguments given above no longer hold,
and where the coupling constants and associated vertex form factors, required
in any meson-exchange model, are difficult to constrain.

In 1993 first evidence for the $\Lambda_c$(2595) resonance was reported
by the CLEO collaboration \cite{CLEO1} and subsequently confirmed by several
other experiments \cite{E6872,ARGUS2,CLEO2}. Nowadays it is generally 
accepted that this resonance is the charmed counterpart of 
the $\Lambda$(1405) \cite{PDG}. In the J\"ulich $\bar KN$
potential model \cite{MG} the latter state is generated dynamically. 
It appears as a $\bar KN$ quasi-bound state and is produced by the strongly attractive 
interaction due to the combined effect of $\omega$, $\rho$ and scalar-meson 
exchanges, which add up coherently in the $\bar KN$ channel. 
When extending the J\"ulich meson-exchange model to the charm sector via SU(4)
symmetry one expects likewise the appearance of a quasi-bound state, namely 
in the $DN$ channel. Thus, one can actually utilize the experimentally known mass 
of the $\Lambda_c$(2595) resonance as an additional constraint for fixing 
parameters of the $DN$ interaction. 

Indeed such a strategy was already followed in recent studies of the 
$\bar KN$ and $DN$ interaction within chiral unitary (and related) approaches,
where likewise the $\Lambda (1405)$ resonance but also states in the $DN$ system
are generated dynamically \cite{Lutz04,Hofmann05,Mizutani06,Lutz06,GarciaRecio09}. 
In those approaches the strong attraction is also provided by vector-meson 
exchange \cite{Hofmann05}, by the Weinberg-Tomazawa (WT) term \cite{Lutz04,Mizutani06,Lutz06}, 
or by an extension of the WT interaction to an SU(8) spin-flavor scheme \cite{GarciaRecio09}. 
In Refs.~\cite{Mizutani06,Lutz06,GarciaRecio09,Tolos04,Tolos08} the authors argued 
that a state occurring in the $S_{01}$ channel of the charm $C = 1$ sector should be 
identified with the $I=0$ resonance $\Lambda_c(2595)$. (Throughout, 
we use  the standard spectroscopical nomenclature $L_{I \, 2J}$, with $L$
denoting the orbital angular momentum, $I$ the isospin and $J$ the total
angular momentum of the two-particle system.)
Note, however, that some of those works differ with regard to the nature of the 
$\Lambda_c(2595)$, i.e. in the dominant meson-baryon component of this state. 
For example, in the SU(8) spin-flavor scheme of Ref.~\cite{GarciaRecio09} the 
$\Lambda_c(2595)$ appears as a $D^*N$ quasi-bound state rather than a $DN$ state.

We take the opportunity to present here also some results of 
the original J\"ulich $\bar KN$ model \cite{MG}. Over the last years
there has been increasing interest in the properties of the $\bar KN$ 
interaction close to the threshold and the near-by $\Lambda (1405)$ 
resonance, resulting in a vast amount of pertinent publications. 
Indeed, there is still a controversy about the actual value for the 
strong-interaction energy shift of kaonic hydrogen and there are 
on-going  debates about issues like a possible double-pole structure 
of the $\Lambda$(1405) or deeply-bound kaonic states \cite{hypx}.
The initial publication of the J\"ulich group \cite{MG} focussed on a 
detailed account of the ingredients of the $\bar KN$ interaction model 
and on the description of scattering data available at that time. 
Results of the J\"ulich model for quantities that are relevant for
the issues mentioned above were not given. This will also be remedied by
the present paper. Specifically, we provide here the $\bar KN$
scattering length and we determine the pole position corresponding
to the $\Lambda (1405)$ resonance. For the latter it turns out that
also the $\bar KN$ model of the J\"ulich group predicts the existence
of two poles in the corresponding $S_{01}$ $\bar KN$ partial wave. 

The paper is organized in the following way: 
In Sect.~2 a brief description of the ingredients of the $DN$ model 
is given. The interaction Lagrangians, which are used to derive the 
meson-baryon potential, are summarized in an Appendix. 
In Sect.~3 selected results for the J\"ulich $\bar KN$ model 
are presented. 
Results for $DN$ scattering are presented in Sect.~4. Besides
the scattering lengths and the pole positions corresponding to
the $\Lambda_c$(2595) resonance, also some predictions for 
total and differential cross sections are given. Furthermore, we
compare our results with those obtained from a $DN$ interaction
that was derived from the leading-order WT contact term 
\cite{Mizutani06}, assuming also SU(4) symmetry. 
Considering those results allows us to explore the model-dependence 
of the predictions for $DN$ scattering observables. 
Sect.~5 is dedicated to the $\Lambda_c$(2595) resonance and 
focusses on the consequences of the fact that the position of
this resonance coincides practically with the $\pi\Sigma_c$
threshold. In particular, we present results for the $\pi\Sigma_c$ 
invariant mass spectrum which allows us to 
discuss the subtle effects of the slightly different thresholds of the 
$\pi^+\Sigma_c^0$, $\pi^0\Sigma_c^+$, and $\pi^-\Sigma_c^{++}$ 
channels on the various invariant mass distributions. 
Also here a comparison with results based on the SU(4) WT approach 
is made. 
Furthermore, we discuss similarities and differences between the 
$\Lambda$(1405) and the $\Lambda_c$(2595) as dynamically generated states.
The paper ends with a short summary.

\section{Coupled-channel {\boldmath$DN$} model in the meson-exchange framework}
\label{sec:model}

\begin{figure}[t]
\vspace*{+1mm}
\centerline{\psfig{file=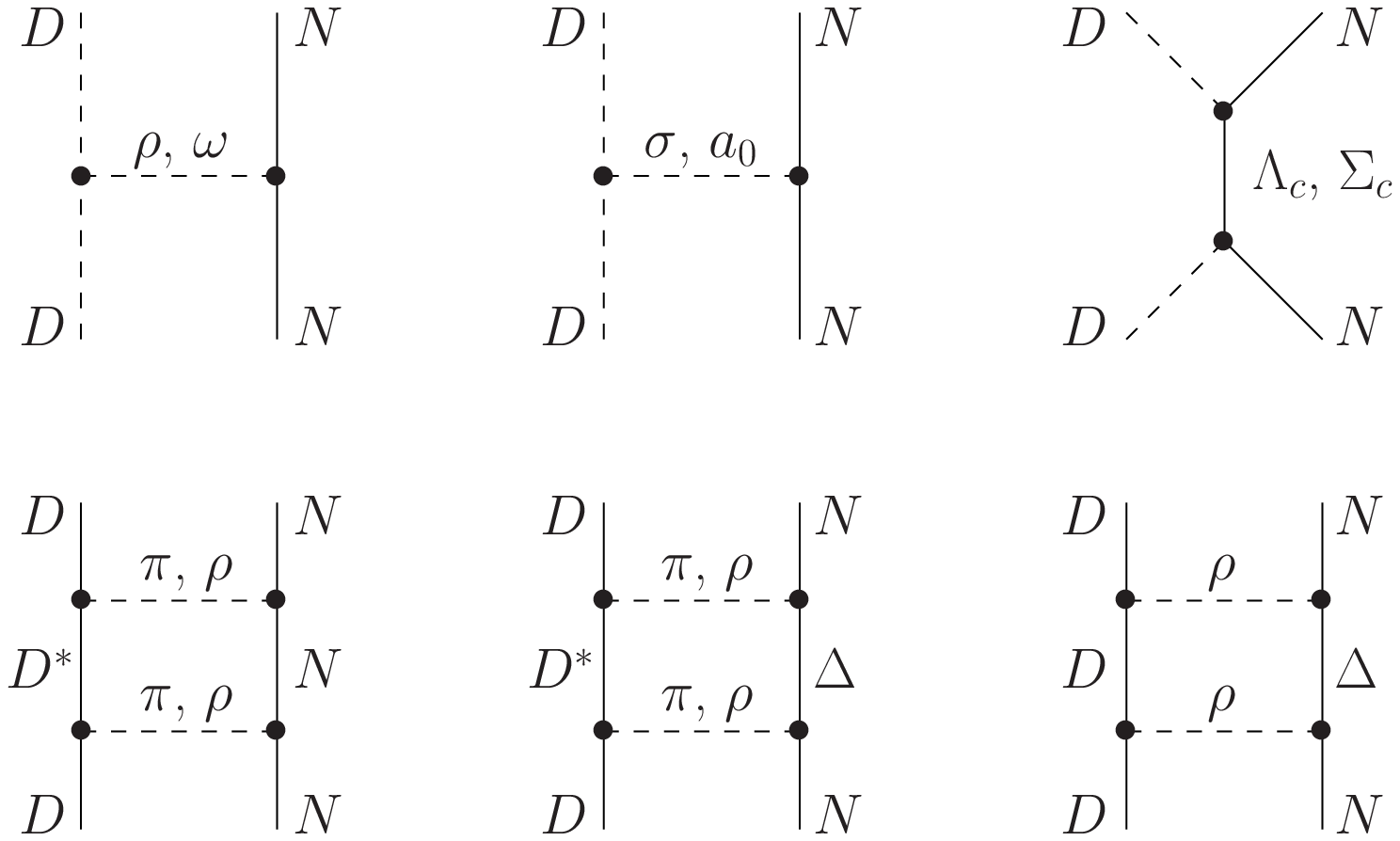,width=11.0cm,height=13.0cm}}
\vspace*{-7.5cm}
\caption{Meson-exchange contributions included in the direct $D N$
interaction.
}
\label{Diadn}
\end{figure}

The $DN$ interaction employed in the present study is constructed
in close analogy to the meson-exchange $\bar K N$ model of the J\"ulich group
\cite{MG} utilizing SU(4) symmetry constraints,
as well as by exploiting the close connection between the $DN$ and $\bar DN$
systems due to G-parity conservation. 
Specifically, we use the latter
constraint to fix the contributions to the direct $DN$ interaction
potential while the former one provides the transitions to and interactions 
in channels that can couple to the $DN$ system.
The main ingredients of the $DN\to DN$ interaction are provided by vector meson
($\rho$, $\omega$) exchange and higher-order box diagrams involving ${D}^*N$,
$D \Delta$, and ${D}^*\Delta$ intermediate states, 
see Fig.~\ref{Diadn}. 

The original $\bar KN$ and $KN$ models of the J\"ulich group include 
besides the exchange of the standard mesons also an
additional phenomenological (extremely short-ranged) repulsive
contribution, a ``$\sigma_{\rm rep}$'', with a mass of about 1.2~GeV 
\cite{MG,Juel2}. This contribution was introduced ad-hoc in
order to achieve a simultaneous description of the empirical $KN$
$S$- and $P$-wave phase shifts \cite{Juel1,Juel2}. Evidently, due to its 
phenomenological nature it remains unclear how that contribution should be 
treated when going over to the $\bar DN$ and $DN$ systems.
This difficulty was circumvented in the construction of the 
$\bar DN$ interaction \cite{Haiden07} by resorting to a recent study of 
the $KN$ interaction \cite{HHK} which provided evidence that a significant 
part of that short-ranged repulsion 
required in the original J\"ulich model could be due to genuine 
quark-gluon exchange processes. Indeed, considering such
quark-gluon mechanisms together with conventional $a_0$(980) meson
exchange instead of that from ``$\sigma_{\rm rep}$'' a comparable if not
superior description of $KN$ scattering could be achieved \cite{HHK}.
From this model the $\bar DN$ interaction \cite{Haiden07} could be derived 
in a straightforward way under the assumption of SU(4) symmetry. 
Furthermore, the extension to the $DN$ system is possible too.
Note, however, that the short-ranged quark-gluon processes, that 
contribute to the $\bar DN$ interaction \cite{Haiden07}, are absent here 
because the quark-exchange mechanism cannot occur in the $DN$ 
interaction due to the different quark structure of the $D$-meson
as compared to $\bar D$.
 
\begin{figure}[t]
\vspace*{+1mm}
\centerline{\psfig{file=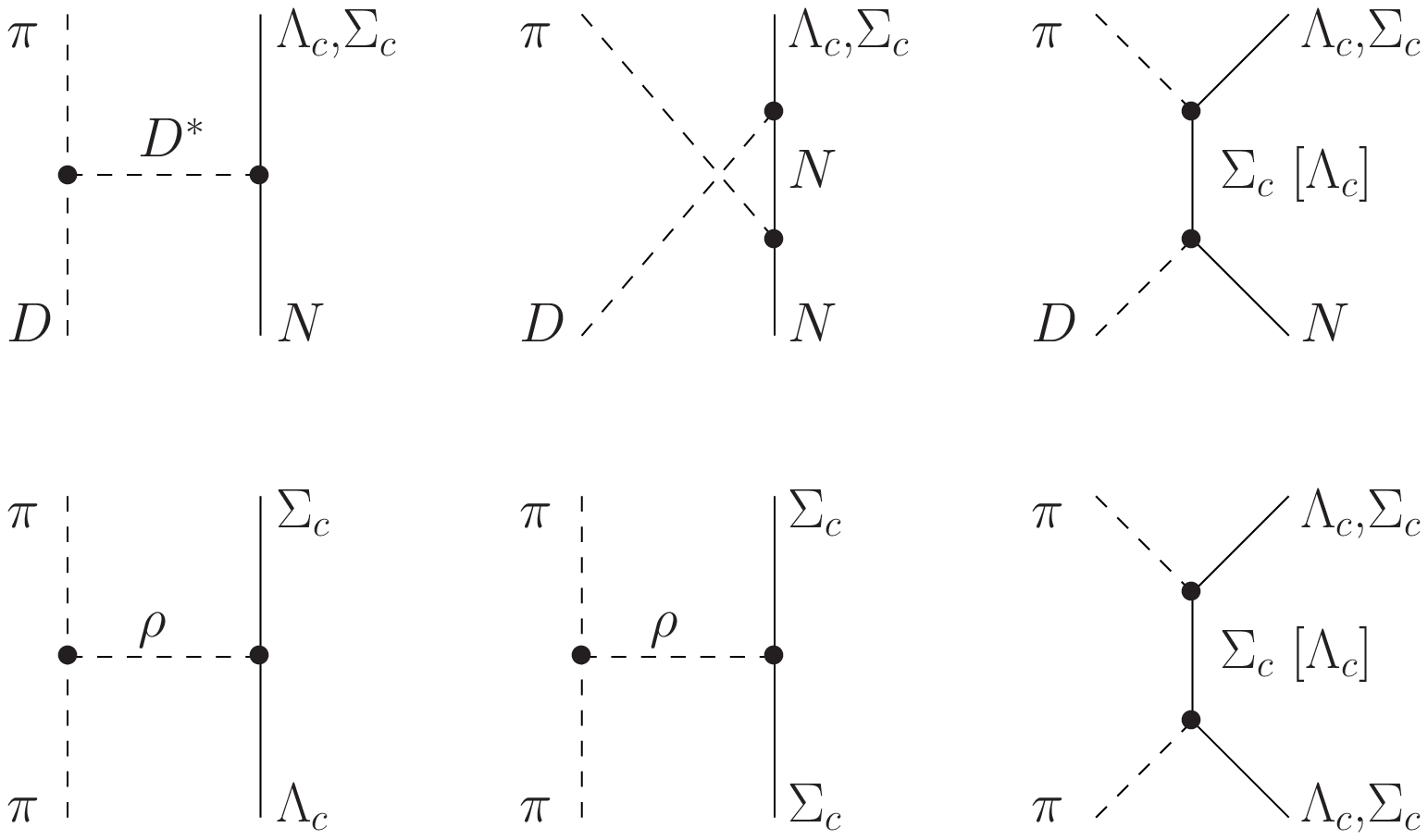,width=11.0cm,height=13.0cm}}
\vspace*{-7.5cm}
\caption{Meson-exchange contributions included in the 
$DN \rightarrow \pi\Lambda_c, \pi\Sigma_c$ transition potentials
and in the 
$\pi\Lambda_c, \pi\Sigma_c \rightarrow \pi\Lambda_c, \pi\Sigma_c$
interactions.
}
\label{Dialp}
\end{figure}

As far as the coupling to other channels is concerned, we follow here the
arguments of Ref.~\cite{MG} and  take into account only the channels
$\pi\Lambda_c(2285)$ and $\pi\Sigma_c(2455)$. Furthermore, we restrict ourselves
to vector-meson exchange and we do not consider any higher-order diagrams in
those channels. Pole diagrams due to the $\Lambda_c(2285)$ and $\Sigma_c(2455)$
intermediate states are, however, consistently included in all channels.
The various contributions to the $DN \to \pi\Lambda_c, \pi\Sigma_c$ transition 
potentials and to the $\pi\Lambda_c, \pi\Sigma_c \to \pi\Lambda_c, \pi\Sigma_c$
interaction, taken into account in the present model, are depicted in
Fig.~\ref{Dialp}.

The interaction Lagrangians, which are used to derive the meson-baryon 
potentials for the different channels, are summarized in the Appendix. 
Based on the resulting interaction potentials ${\cal V}_{ij}$ 
($i,j=$ $DN$, $\pi\Lambda_c$, $\pi\Sigma_c$)
the corresponding reaction amplitudes ${\cal T}_{ij}$ are
obtained by solving a coupled-channel Lippmann-Schwinger-type 
scattering equation within the framework of time-ordered perturbation theory,
\begin{eqnarray}
{\cal T}_{ij} = {\cal V}_{ij} + \sum_k {\cal V}_{ik} {\cal G}^0_k {\cal T}_{kj} \ ,
\label{LSE}
\end{eqnarray}
from which we calculate the observables in the standard way \cite{Juel1}.

The assumed SU(4) symmetry and the connection with the $\bar KN$ model, respectively, 
allows us to fix most of the parameters - the coupling constants and the 
cut-off masses at the vertex form factors of the occurring meson-meson-meson and
meson-baryon-baryon vertices, cf. Ref.~\cite{MG}. A list with the specific values of 
the pertinent parameters can be found in the Appendix. 

When solving the coupled-channel Lippmann-Schwin\-ger equation with this interaction 
model we observe that two narrow states are generated dynamically below the $DN$ 
threshold, one in the $S_{01}$ partial wave and the other one in the $S_{11}$ 
phase. In view
of the close analogy between our $DN$ model and the corresponding $\bar KN$
interaction \cite{MG} this is not too surprising, because the latter also yields
a quasi-bound state in the $S_{01}$ channel which is associated with the
$\Lambda (1405)$ resonance. The bound states in both  $\bar KN$ and $DN$ are
generated by the strongly attractive interaction due to the combined effect of
$\omega$, $\rho$ and scalar-meson exchanges, which add up coherently in specific
channels.

Following the arguments in Refs.~\cite{Mizutani06,Lutz06,GarciaRecio09,Tolos04,Tolos08} 
we identify the narrow state occurring in the $S_{01}$ channel with the $I=0$ resonance 
$\Lambda_c(2595)$. Furthermore, we identify the state we get in the $S_{11}$ 
channel with the $I=1$ resonance $\Sigma_c(2800)$ \cite{PDG}.
(For a different scenario, see \cite{Dong}.) 
In order to make sure that the $DN$ model incorporates the above features also
quantitatively, the contributions of the scalar mesons to the $DN$
interaction are fine-tuned so that the position of those states generated 
by the model coincide with the values given by the Particle Data 
Group \cite{PDG}. 
This could be achieved by a moderate change in the coupling constants of the 
$\sigma$ meson (from $1$ to $2.6$) and the $a_0$ meson (from $-2.6$ to $-4.8$), 
cf. the Table in the Appendix. 
We would like to stress that anyhow the application of SU(4) symmetry (and even 
SU(3) symmetry) to the scalar-meson sector is problematic, as discussed in 
Ref.~\cite{Haiden07}.
In the present paper we will show results for 
the latter interaction which we consider as our basic model. 
However, we present also $DN$ cross sections based on the parameter set that 
follows directly from the $\bar KN$ and $KN$ studies \cite{Juel1,Juel2,HHK}
by assuming SU(4) symmetry. Those results may be considered as a measure for 
the uncertainty in our model predictions for $D N$. 

With regard to the $D^*$(2010) exchange that contributes to the 
$DN\to \pi\Lambda_c$ and $DN\to \pi\Sigma_c$ transition potentials (cf. Fig.
\ref{Dialp}) it should be said that the corresponding form factors cannot 
be inferred from the $\bar KN$ model \cite{MG} via SU(4) arguments. We
fixed the relevant cut-off masses somewhat arbitrarily to be about 1 GeV 
larger than the exchange mass. Anyway, variations in those cut-off masses
have very little influence on the results in the $DN$ channel. Moreover,
within the spirit of the basic model such variations can be easily compensated
by re-adjusting the coupling constants in the scalar sector so that again
the masses of the $\Lambda_c(2595)$ and $\Sigma_c(2800)$ resonances are
reproduced, as discussed above. But the width of those states are certainly
sensitive to the values used for those cut-off masses. 

\section{Selected results of the J\"ulich {\boldmath$\bar K N$} model}

Before discussing the results of the J\"ulich $DN$ model in detail, let
us first present some selected results of the corresponding $\bar K N$ 
model. The general scheme and also the explicit expressions for the 
various contributions to the $\bar KN$ interaction potential of the 
J\"ulich group are described in detail in Ref.~\cite{MG}.
In the original publication of the model, results for total cross 
sections of the reaction channels
$K^-p \to K^-p$, $K^-p \to K^0n$,
$K^-p \to \pi^0\Lambda$, $K^-p \to \pi^0\Sigma^0$,
$K^-p \to \pi^-\Sigma^+$, and $K^-p \to \pi^+\Sigma^-$
were presented and compared with available data.
One can see \cite{MG} that the model yields a quite satisfactory
description of the available experimental information 
up to $\bar K$ laboratory momenta of $p_{lab} \approx$ 300 MeV/c. 
In the present paper we refrain from showing those results
again but we want to focus on additional predictions of the
model that have not been included in Ref.~\cite{MG}. 
First this concerns the behaviour close to the $\bar K N$
threshold. The original J\"ulich potential is derived under the 
assumption of isospin symmetry and the reaction amplitudes were 
obtained by solving a Lippmann-Schwinger-type scattering equation 
(\ref{LSE})
in the isospin basis using averaged masses of the involved
baryons and mesons. The corresponding $S$-wave scattering lengths 
$a$ for $I=0$, $I=1$, and for $K^-p$ ($a_{K^-p} = (a_0+a_1)/2$) 
are summarized in Table~\ref{tab1}. 
In order to enable a detailed comparison with available 
empirical information in the threshold region,
now we also performed a calculation in the particle basis. 
Using the physical masses of the baryons and mesons
allows us to take into account the isospin-breaking
effects generated by the known mass splittings within the
involved isospin multiplets, and specifically between
the $K^-$ and $\bar K^0$ masses. 
The $K^-p$ scattering length calculated in this way is also
given in Table~\ref{tab1}. It differs significantly from
the one obtained in the isospin basis (labelled by (I)).

\begin{table}[t]
\renewcommand{\arraystretch}{1.3}
\caption{$\bar K N$ results of the J\"ulich model \cite{MG}.
The value for the $K^-p$ scattering length obtained in the 
isospin-symmetric calculation is marked with (I).
In case of kaonic hydrogen we present the strong-interaction
energy shift $\Delta E$ and the width $\Gamma$ of the $1s$
level. The result for the J\"ulich model is obtained from
the $K^-p$ scattering length with the modified Deser-Trueman
formula \cite{MRR}. 
}
\label{tab1}
\centering
\begin{tabular}{|l|c|c|}
\hline
 & J\"ulich model \cite{MG} & experiment \cr
\hline
 \multicolumn{3}{|c|}{scattering lengths [fm]}\\
\hline
$a_{I=0}$ & $-$1.21 + $i$1.18 & \cr 
$a_{I=1}$ &  1.01 + $i$0.73 & \cr 
$a_{K^-p}$ (I) & $-$0.10 + $i$0.96 & \cr 
\hline
$a_{K^-p}$ & $-$0.36 + $i$1.15 & \cr 
\hline
 \multicolumn{3}{|c|}{kaonic hydrogen}\\
\hline
$\Delta E$ & 217 eV & 323$\pm$63$\pm$11 eV \cite{kek}\cr 
                  &  & 193$\pm$37$\pm$6 eV \cite{dear} \cr 
$\Gamma$ & 849 eV & 407$\pm$208$\pm$100 eV \cite{kek}\cr 
                  &  & 249$\pm$111$\pm$39 eV \cite{dear} \cr 
\hline
 \multicolumn{3}{|c|}{threshold ratios -- Eq.~(\ref{rat})}\\
\hline
$\gamma$  & 2.30 & 2.36$\pm$0.04 \cite{Nowak} \cr 
$R_c$  & 0.65 & 0.664$\pm$0.011 \cite{Nowak} \cr 
$R_n$  & 0.22 & 0.189$\pm$0.015 \cite{Nowak} \cr 
\hline
 \multicolumn{3}{|c|}{pole positions [MeV]}\\
\hline
$S_{01}$ & 1435.8 + $i$25.6 & \cr 
$S_{01}$ & 1334.3 + $i$62.3 &\cr 
\hline
\end{tabular}
\end{table}

Recently a thorough study of the $K^-p$ scattering length and its theoretical 
uncertainties within chiral SU(3) unitary approaches was presented \cite{Bora}. 
The full calculation which included the WT contact interaction
at leading chiral order, the direct and crossed Born terms as well as
contact interactions from the Lagrangian of second chiral order yielded a
scattering length of $a_{K^-p} = -1.05 + {\rm i} 0.75\, {\rm fm}$.
Obviously, the result obtained for the J\"ulich meson-ex\-chan\-ge model differs from this
value, but it is still within the $1\sigma$ confidence region  given in 
Ref.~\cite{Bora}.
 
Results for the strong-interaction energy shift $\Delta E$ and the width 
$\Gamma$ of the $1s$ level of kaonic hydrogen, deduced from
the $K^-p$ scattering length with the modified Deser-Trueman
formula \cite{MRR}, are also given in Table~\ref{tab1}. Interestingly,
the prediction of the J\"ulich model for $\Delta E$ agrees well with the
DEAR result \cite{dear}  while $\Gamma$ is roughly in line with the value found in the KEK experiment \cite{kek}. 
Note, however, that the experimental results for kaonic hydrogen were not available at the
time when the J\"ulich $\bar K N$ model was constructed and, therefore, not included in
the fitting procedure.  
Since the KEK and DEAR results are not compatible with each other it is important to resolve 
this discrepancy between the experimental results. Values of higher level of precision are
expected to be reached by the ongoing SIDDHARTA experiment at LNF \cite{siddharta}.

\begin{figure*}[t]
\begin{center}
\includegraphics[height=100mm]{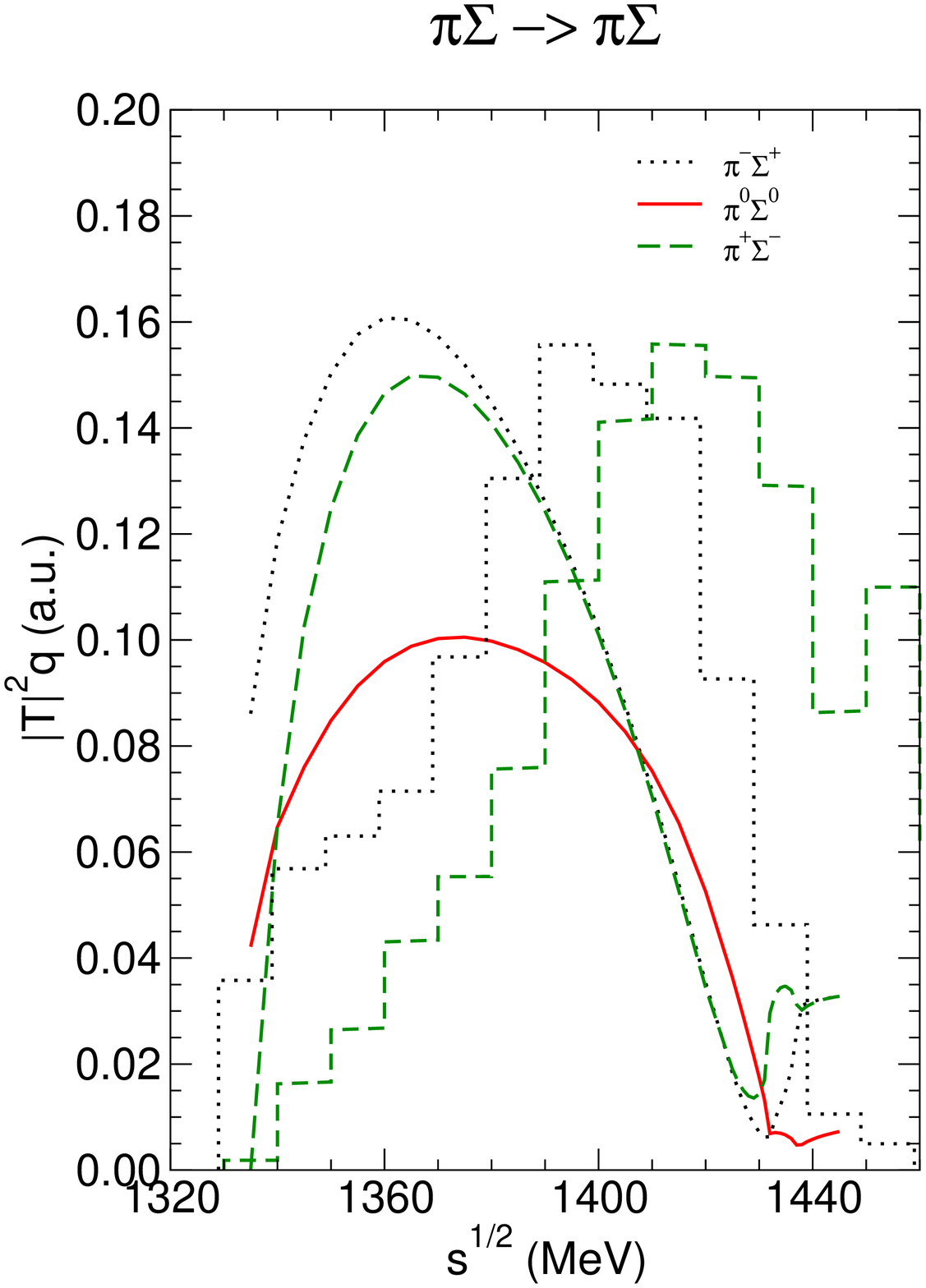}
\includegraphics[height=100mm]{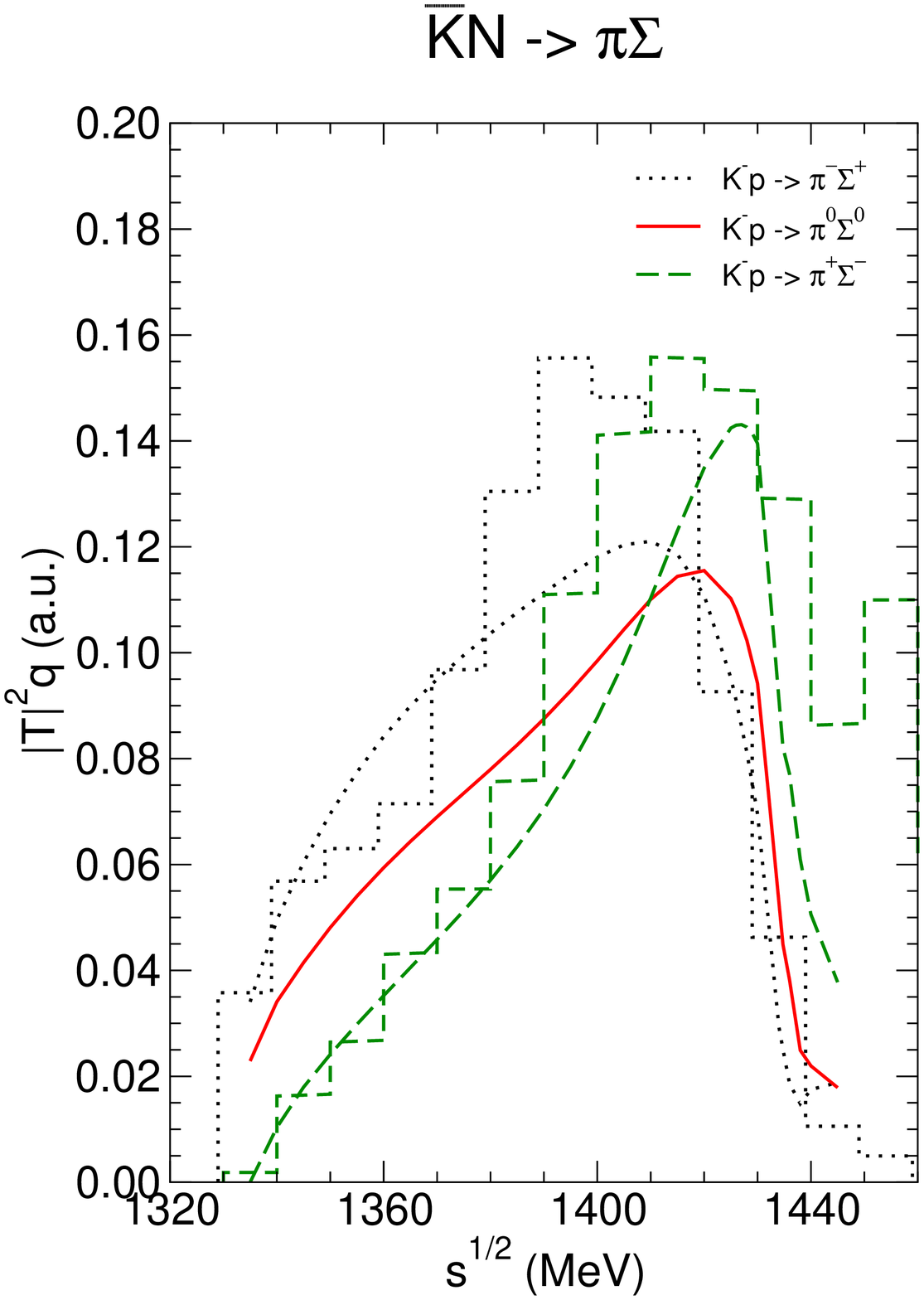}
\caption{$\pi\Sigma$ invariant mass spectrum.
Left are results based on the $\pi\Sigma \to \pi\Sigma$ 
$T$-matrix and right the ones for $K^-p\to \pi\Sigma$. 
The histograms are experimental results for the $\pi^-\Sigma^+$
(dotted) and $\pi^+\Sigma^-$ (dashed) channels taken from 
Ref.~\cite{Hem}. 
}
\label{fig:2}
\end{center}
\end{figure*}

Of interest are also the three measured threshold ratios~\cite{Nowak} of 
the $K^-p$ system, which are defined by
\begin{eqnarray} 
\label{rat}
\gamma&=&\frac{\Gamma(K^-p \rightarrow \pi^+\Sigma^-)}{\Gamma(K^-p \rightarrow
\pi^- \Sigma^+)}, \nonumber \\
R_c&=&\frac{\Gamma(K^-p\rightarrow \hbox{charged particles})}{\Gamma(K^-p
\rightarrow \hbox{all})}, \nonumber \\
R_n&=&\frac{\Gamma(K^-p \rightarrow \pi^0 \Lambda)}{\Gamma(K^-p \rightarrow
\hbox{all neutral states})}.
\end{eqnarray} 
The corresponding predictions of the J\"ulich model are given in
Table~\ref{tab1}. As one can see, the values are remarkably close to
the experimental ones, specifically when one keeps in mind that these
threshold ratios were not included in the fitting procedure when the
model was constructed. 

Re-calculating total cross sections in the particle basis yielded results
that do not differ very much from those shown in Ref.~\cite{MG}
(obtained in isospin basis) for momenta $p_{lab} \gsim$~100 MeV/c where 
data are available. In particular, the changes are small in comparison to 
the given experimental error bars. Thus, we do not present the corresponding 
results here. 

Let us now come to the $\Lambda$(1405). As already said above, also
in the J\"ulich model this structure is generated dynamically.
It is predicted ``unambiguously''\cite{MG} to be a quasi-bound
$\bar KN$ state. When searching in the region between the 
$\pi\Sigma$ and $\bar KN$ threshold we find two poles in the
complex plane for the relevant $S_{01}$ partial wave. 
The values of the pole positions are listed in Table~\ref{tab1}.
In view of the extensive discussion of the two-pole structure of
the $\Lambda$(1405) over the last ten years or so 
\cite{Bora,Oller01,Carmen02,Jido:2003cb,Oller05,Borasoy:2005ie,Revai09}
it is certainly
not surprising that the J\"ulich meson-exchange potential of the
$\bar KN$ interaction generates two such poles as well. 
But back in
1989 when the model was constructed this was not an issue yet. Thus,
no attempt was made to examine the pole structure of the amplitude
in detail and, therefore, this feature remained undiscovered. 

The pole structure predicted by the J\"ulich model turns out to be
quantitatively very similar to the one of the full result of 
the chiral SU(3) unitary approach by Borasoy et al. \cite{Bora}.
One pole, the $\bar KN$ ``bound state'', is located fairly close
to the $\bar KN$ threshold and to the physical real axis
while the other one is close to the $\pi\Sigma$ threshold 
and has a significantly larger imaginary part. Comparable 
results were also reported in Ref. \cite{Borasoy:2005ie}.
As emphasized in Refs.~\cite{Bora,Borasoy:2005ie}, 
this structure differs qualitatively from the scenario where
only the leading-order WT contact interaction is taken into account 
\cite{Oller01,Carmen02,Jido:2003cb,Oller05,OR98} which suggests a 
very pronounced two-pole structure of the $\Lambda$(1405). 
For example, the interaction used in Refs.~\cite{Oller01,Oller05} 
leads to two poles that are both very close to the physical region. 
In case of the WT models of Refs.~\cite{Carmen02,Jido:2003cb,OR98} the second 
pole shows a larger width,
more in line with our results and those of Refs.~\cite{Bora,Borasoy:2005ie},
though the two poles are still fairly close together. 

In Ref.~\cite{MG} results for the quantity $|T_{\pi\Sigma}|^2\cdot\,q$ 
were presented, which is commonly associated and compared with the
$\pi\Sigma$ mass distributions, i.e.
\begin{equation}
\frac{d\sigma}{dm_{\pi\Sigma}} \propto |T_{\pi\Sigma}|^2\,q \ .
\label{INV}
\end{equation}
Here, $q$ is the center-of-mass momentum of the $\pi\Sigma$ system.
Measurements of the $\pi\Sigma$ invariant mass distribution for reactions 
like $\bar KN \to \pi\pi\pi\Sigma$ \cite{Hem}  or 
$\pi N \to K \pi\Sigma$ \cite{Tho} provide the main
experimental evidence for the $\Lambda$(1405) resonance. 
 As we realize now, the results published in Ref.~\cite{MG} 
were not correct, because of an error in the phase-space factor in
the computer code. Moreover, in this reference only the contribution
from  $I=0$ alone was considered. 

In Fig.~\ref{fig:2} we present results for different
charge channels ($\pi^-\Sigma^+$, $\pi^0\Sigma^0$, $\pi^+\Sigma^-$)  
in the final state and consider $\pi\Sigma \to \pi\Sigma$ as
well as $\bar KN \to \pi\Sigma$ transitions, and compare them with 
the $\pi\Sigma$ mass distribution measured in the reaction 
$K^- p \to \pi\pi\pi\Sigma$ \cite{Hem}. We display here curves 
for the individual $T$-matrices because we want to illustrate
the differences between the various amplitudes. 
Note that the ``true'' amplitude to be inserted in Eq.~(\ref{INV}) 
will be a coherent sum of transition amplitudes from all allowed
intermediate states to a specific final state, weighted with 
coefficients that reflect the details of the reaction mechanism
\cite{Oller01}. 

As already observed by others in the past \cite{Jido:2003cb,Jido09}, 
there
is a remarkable difference between the invariant mass spectrum
due to the $\pi\Sigma \to \pi\Sigma$ amplitude (left-hand side)
and the one due to the $\bar KN \to \pi\Sigma$ amplitude 
(right-hand side). This is due 
to the fact that the two poles in the $I=0$ $S$-wave amplitude 
have different widths and couple also differently to the $\bar K N$ and 
$\pi \Sigma$ channels ~\cite{Jido:2003cb}. 
In case of the J\"ulich potential the former
shows little resemblance with the $\pi\Sigma$ invariant mass
spectrum given in Ref.~\cite{Hem} while the results based
on the $\bar KN \to \pi\Sigma$ amplitude are roughly in line
with the empirical information. 

Note that there are also significant differences in the
invariant mass distributions due to the $K^-p \to \pi\Sigma$ 
amplitudes for the different possible charge states of $\pi\Sigma$.
Specifically, the peak positions for $\pi^-\Sigma^+$ and $\pi^+\Sigma^-$ 
differ by almost 30 MeV. Interestingly, the experimental
invariant mass distributions for the two charge states, cf. the dotted
and dashed histograms in Fig.~\ref{fig:2}, respectively, 
seem to show a similar separation of their peak and both agree 
roughly with the
corresponding predictions of the J\"ulich $\bar KN$ potential. 
We should mention, however, that the experimental $\pi^+\Sigma^-$
mass spectrum is afflicted by fairly large background
contributions \cite{Hem}. 

These differences in the invariant mass distributions are caused 
primarily by the interference between the $I=0$ and $I=1$ amplitudes 
\cite{Nacher:1998mi} when evaluating the observables in the particle 
basis. They are not due to isospin-breaking effects. Indeed, we found
that an isospin-symmetric calculation based on averaged masses yields 
very similar results. Note that then the 
$\pi^-\Sigma^+ \to \pi^-\Sigma^+$ and $\pi^+\Sigma^- \to \pi^+\Sigma^-$ 
results would coincide. 

In this context let us mention that the interest in the lineshape of the 
$\Lambda(1405)$ has been recently revived with the $\Lambda(1405)$ 
photoproduction experiment on a proton target at CLAS$@$JLAB. 
This experiment is 
providing the first results on the $\Lambda(1405)$ photoproduction  
analyzing all three $\pi \Sigma$ decay modes. Preliminary results show 
differences in the $\Lambda(1405)$ lineshapes as well as different 
$\Lambda(1405)$ differential cross sections for each $\pi \Sigma$ 
decay mode \cite{schumacher}.

\begin{table}
\renewcommand{\arraystretch}{1.3}
\caption{$D N$ results of the meson-exchange model and the
SU(4) WT model of Ref.~\cite{Mizutani06}. The pole positions
of the latter model were published in \cite{GarciaRecio09}. 
}
\label{tab2}
\centering
\begin{tabular}{|l|c|c|}
\hline
 & meson-exchange model & SU(4) $DN$ model \cite{Mizutani06} \cr
\hline
 \multicolumn{3}{|c|}{scattering lengths [fm]}\\
\hline
$a_{I=0}$ & $-$0.41 + $i$\,0.04 & $-$0.57 + $i$\,0.001 \cr 
$a_{I=1}$ & $-$2.07 + $i$\,0.57 & $-$1.47 + $i$\,0.65 \cr 
\hline
 \multicolumn{3}{|c|}{pole positions [MeV]}\\
\hline
$S_{01}$ & 2593.9 + $i$\,2.88 &  2595.4 + $i$\,1.0  \cr
$S_{01}$ & 2603.2 + $i$\,63.1 &  2625.4 + $i$\,51.5 \cr
$S_{01}$ &                    &  2799.5 + $i$\,0.0  \cr
$S_{11}$ & 2797.3 + $i$\,5.86 &  2661.2 + $i$\,18.2 \cr
$S_{11}$ &                    &  2694.7 + $i$\,76.5 \cr
$P_{01}$ & 2804.4 + $i$\,2.04 & \cr
\hline
\end{tabular}
\end{table}
\section{Results for {\boldmath$D N$}}

As already stated above, the SU(4) extension of the J\"ulich $\bar KN$
model to the $DN$ interaction generates narrow states in the $S_{01}$
and $S_{11}$ partial waves which we identify with the experimentally 
observed $\Lambda_c(2595)$ and $\Sigma_c(2800)$ resonances, 
respectively. Not surprisingly, we find an additional pole in the 
$S_{01}$ partial wave, located close to the other one. Similar to the 
corresponding state in the $\bar KN$ sector, the second pole has a 
much larger width, cf. Table~\ref{tab2}. 
Interestingly, our model even generates a further state, name\-ly in 
the $P_{01}$ partial wave, which, after  fine-tuning (cf. Sect.~2), 
lies at 2804 MeV, i.e. just below the $DN$ threshold.  We are tempted to 
identify this state with the $\Lambda_c(2765)$ resonance, whose quantum 
numbers are not yet established \cite{PDG}. 
Though we do not reproduce the resonance energy quantitatively, we
believe that further refinements in the $DN$ model, specifically the inclusion
of the $\pi\pi\Lambda_c$ channel in terms of an effective $\sigma\Lambda_c$
channel, can provide sufficient additional attraction for obtaining also
quantitative agreement. The mechanism could be the same as in the case of the Roper
($N^*(1440)$) resonance, which is generated dynamically in the J\"ulich $\pi N$
model \cite{Krehl,Achot}. Here the required strong attraction is produced via
the coupling of the $\pi N$ $P$-wave (where the Roper occurs) to the $S$-wave in
the $\sigma N$ system, facilitated by the different parities of the $\pi$ and
$\sigma$ mesons. Besides shifting the resonance position, the coupling to 
an effective $\sigma\Lambda_c$ would certainly also increase the width 
significantly, as is required for a reproduction of the experimental 
information \cite{PDG}.

\begin{figure}[t]
\begin{center}
\includegraphics[height=90mm,angle=-90]{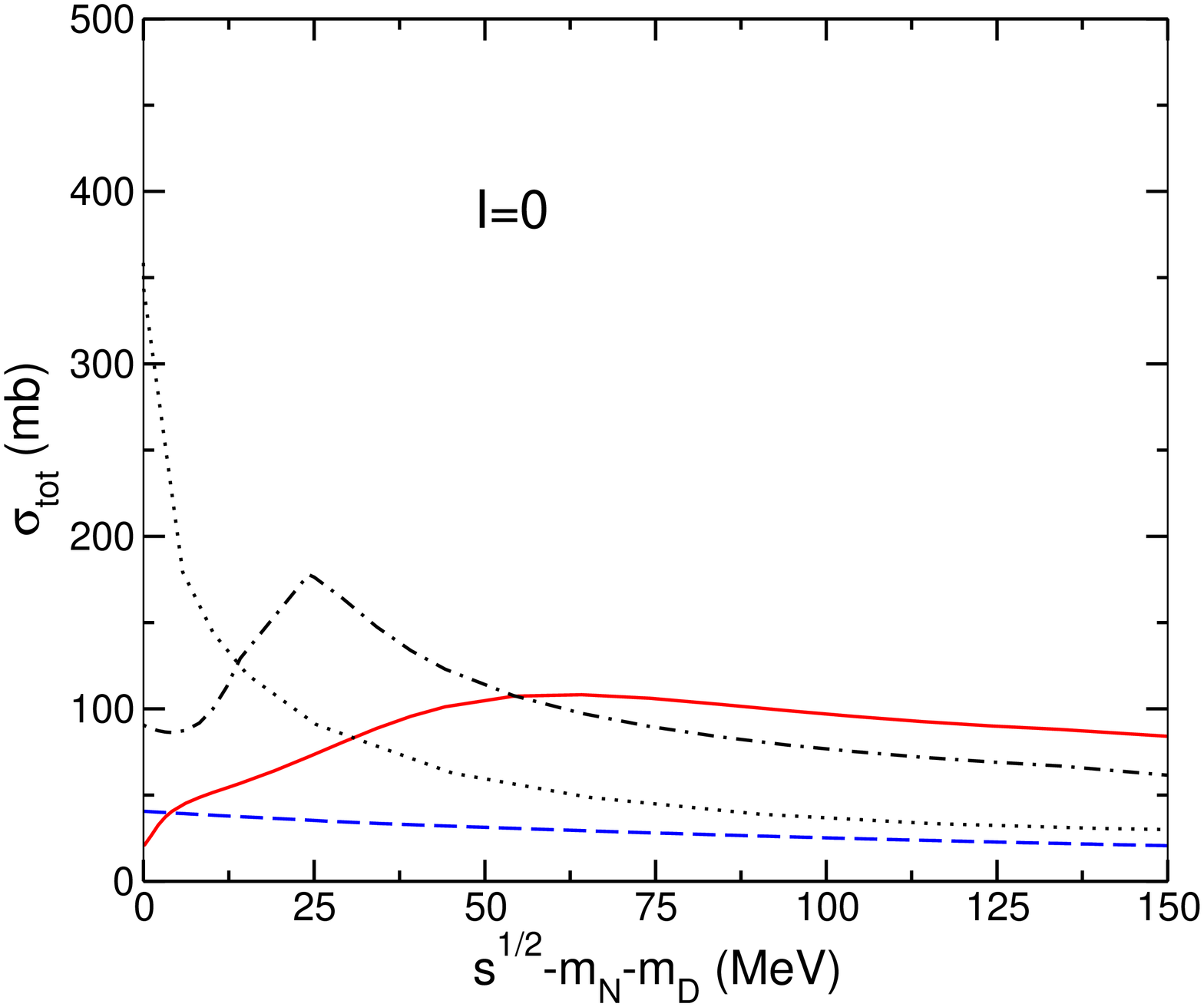}\\
\includegraphics[height=90mm,angle=-90]{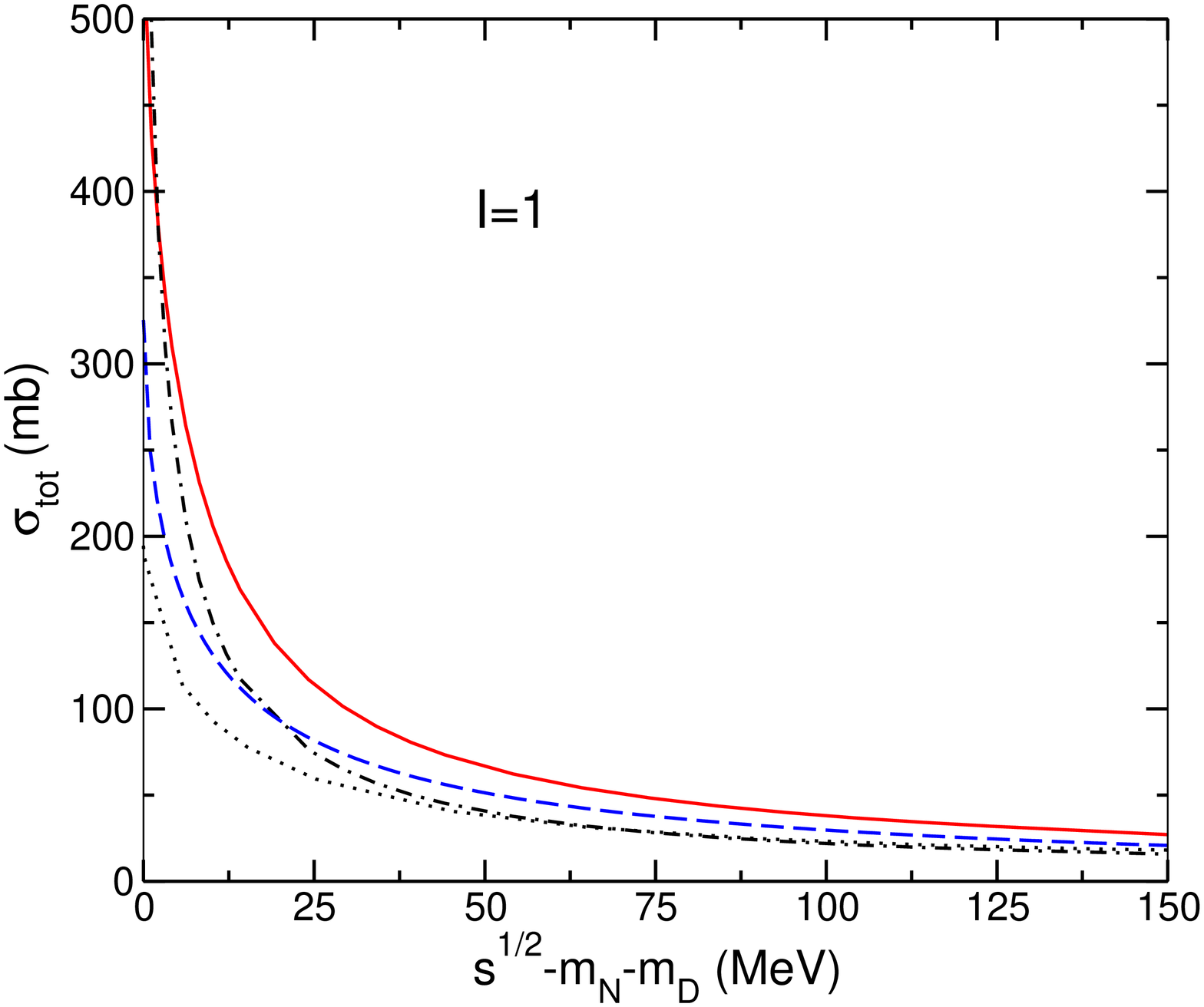}
\caption{$DN$ cross sections for the isospin
$I=0$ (top) and $I=1$ (bottom) channels as a function of 
$\epsilon = \sqrt{s} - m_N - m_{D)}$. 
The solid lines are the results of our meson-exchange model.
The dash-dotted lines are based on parameters taken directly
from earlier $\bar K N$ and $K N$ potentials \cite{Juel1,Juel2,HHK},
cf. text, while 
the dashed lines are predictions of the SU(4) WT model of Ref.~\cite{Mizutani06}.
The dotted lines are corresponding results for $\bar KN$ of the
J\"ulich meson-exchange model \cite{MG}. 
}
\label{figdn}
\end{center}
\end{figure}

For comparison we include here some predictions of the $DN$ interaction
presented in Ref.~\cite{Mizutani06} which is based on the leading-order WT
contact term.  
The $DN$ interaction of Ref.~\cite{Mizutani06} has also been adjusted to reproduce 
the $\Lambda_c(2595)$ resonance. However, compared to the J\"ulich 
meson-ex\-chan\-ge model, the total number of poles and their energies are 
different, cf. Table~\ref{tab2}.
Three $S_{01}$ and two $S_{11}$ states appear up to 2800 MeV in the SU(4) $DN$ WT model, 
as reported in Ref.~\cite{GarciaRecio09} in the analysis of the SU(4) sector. 
Among them, there is a $S_{11}$ state at $2694 \ {\rm MeV}$ with a width of 
$\Gamma= 153 \ {\rm MeV}$ that strongly couples to the $DN$ channel. 
Thus, it is observed as a structure in the real axis close to the $DN$ threshold with 
similar effects as the $S_{11}$ resonance at $2797$~MeV of our model. 

\begin{figure}
\begin{center}
\includegraphics[height=90mm,angle=-90]{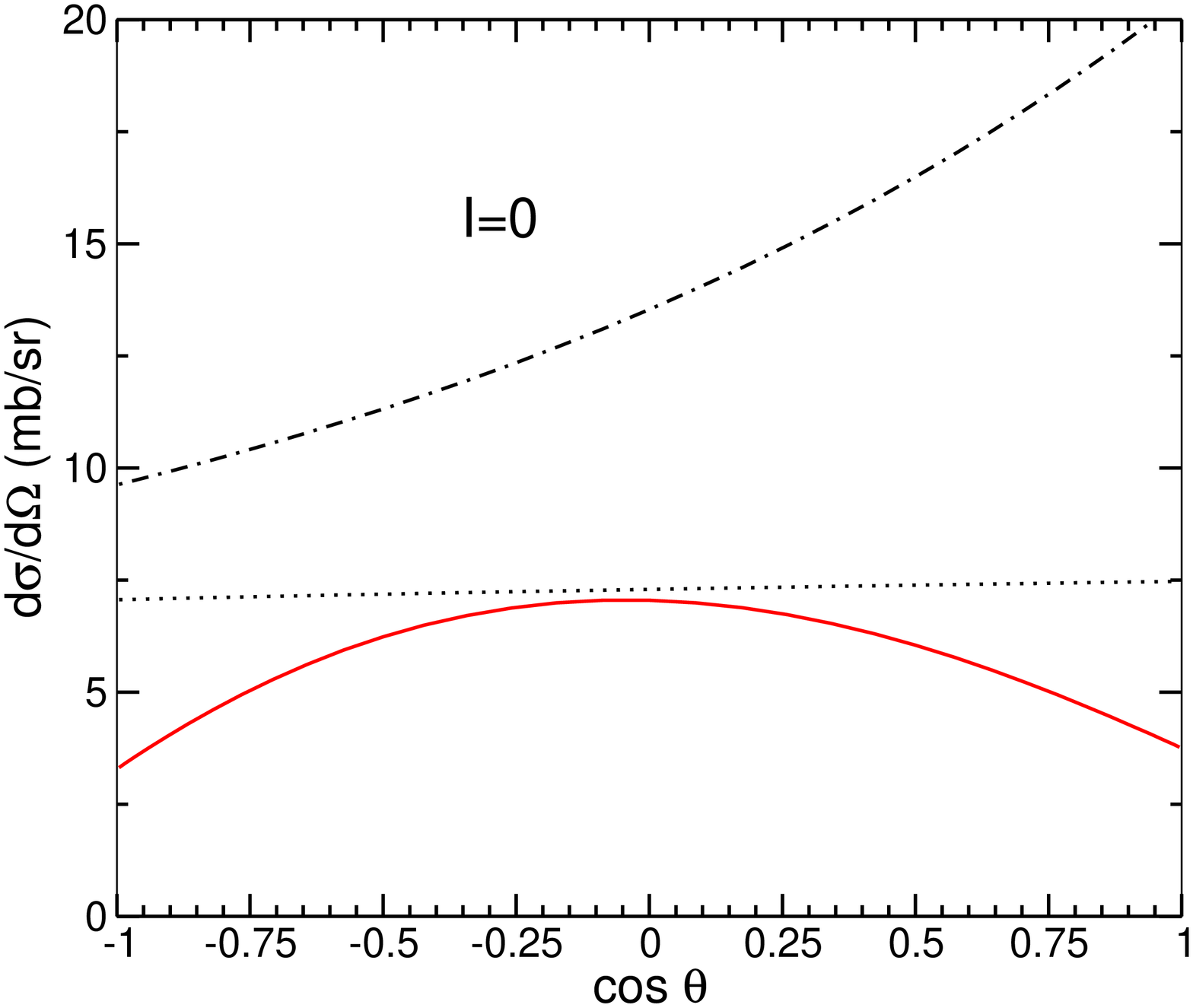}\\
\includegraphics[height=90mm,angle=-90]{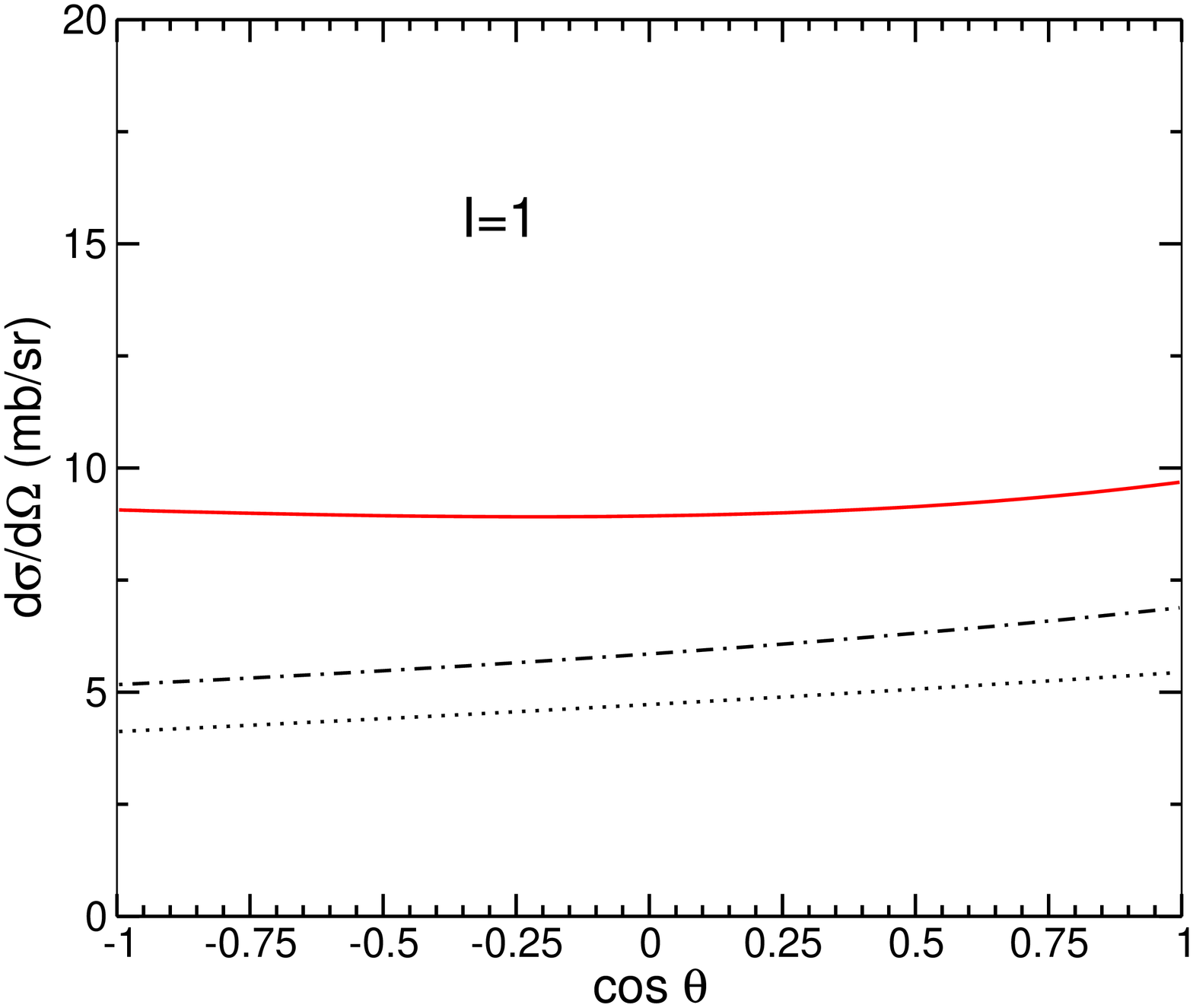}
\caption{Differential cross sections for $DN$ and $\bar KN$ for the isospin
$I=0$ (top) and $I=1$ (bottom) channels at $\epsilon = 25$ MeV. 
Same description as in Fig.~\ref{figdn}. 
}
\label{figdnd}
\end{center}
\end{figure}

Some results for $DN$ scattering are presented in Figs.~\ref{figdn}
and \ref{figdnd}. Specifically, we
show $DN$ cross sections for $I=0$ and $I=1$ based on 
the parameter set that reproduces the positions of the $\Lambda_c(2595)$ 
and $\Sigma_c(2800)$ of the Particle Data Group (solid line) together 
with results that follow directly from the $\bar K N$ and $K N$ studies 
of Refs.~\cite{Juel1,Juel2,HHK} by assuming SU(4) symmetry (dash-dotted line). 
The $DN$ cross sections of the 
SU(4) WT model of Ref.~\cite{Mizutani06} are also displayed (dashed line).  
Finally, we include the $\bar K N$ cross sections of the original J\"ulich
model \cite{MG} for reference (dotted line).

Obviously, the $DN$ cross sections show a significant momentum dependence. 
Furthermore, they are substantially larger
than those we obtain for $\bar DN$ \cite{Haiden07}. 
In particular, the cross section in the isospin $I=1$ channel
amounts to almost 600~mb at threshold. This is not too
surprising in view of the near-by $S_{11}$ quasi-bound state. 
The structure of the cross section in the isospin $I=0$ channel 
is strongly influenced by the $P_{01}$ partial wave where our model
produces a near-threshold quasi-bound state or, in case of the
parameter set directly fixed by SU(4) constraints, a resonance state
around 25 MeV above threshold. 
The $DN$ cross section of the SU(4) WT approach of Ref.~\cite{Mizutani06} 
shows a similar behaviour as the one of the J\"ulich model for the $I=1$ channel. 
As already discussed above, this model generates likewise poles in the $S_{11}$ 
partial wave, though not as close to the $DN$ threshold as the J\"ulich model. 
In case of the $I=0$ channel there are pronounced differences. But this is
not surprising because the SU(4) WT model yields only $S$-wave contributions
while the results of the J\"ulich model are dominated by the $P$-wave.

The quasi-bound state in the $S_{11}$ channel is also
reflected in the scattering lengths, cf. Table~\ref{tab2}, 
namely by the rather large value of the real part of $a_{I=1}$. 
The same situation is observed in the WT approach of \cite{Mizutani06}. In fact, the $S$-wave
scattering lengths predicted by our model and by the WT approach turn out to be 
very similar qualitatively for the $I=1$ as well as for the $I=0$ channel, 
as can be seen in Table~\ref{tab2}. 
For completeness, let us mention here that the scattering lengths of the $DN$ interaction 
of Hofmann and Lutz \cite{Hofmann05}, reported in Ref.~\cite{Lutz06},
amount to about $-0.4$~fm for both isospin channels. In agreement with that work we
find that the imaginary part in the J\"ulich model is negligibly small for $I=0$. 
However, the imaginary part in  the  $I=1$ channel for our $DN$ model is not negligible, 
contrary to Ref.~\cite{Lutz06}. This is due to the fact that in Ref.~\cite{Lutz06} there is no 
quasi-bound state close to $DN$ threshold but lies 180~MeV below.

Angular distributions for the reaction $DN \to DN$ are shown in
Fig.~\ref{figdnd}. Obviously, in the $I=0$ case there is a strong anisotropy 
already at fairly low
momenta. It is due to significant contributions in the $P_{01}$ partial wave in
this momentum region induced by the near-threshold quasi-bound state or resonance,
respectively, produced by our model, as discussed above. 
For higher momenta, the differential cross section
becomes forward peaked, similar to the predictions of our model for the $\bar DN$
system \cite{Haiden07}.

Predictions for $DN$ scattering observables in the particle basis
($D^0n \to D^0n$, $D^0p \to D^0p$, $D^0p \to D^+n$) can be found in
Ref.~\cite{Haiden08}.

\section{Discussion of the {\boldmath$\Lambda_c$(2595)}}

The excited charmed baryon  $\Lambda_c$(2595) was first observed by 
the CLEO collaboration \cite{CLEO1} and its existence was later confirmed
in experiments by the E687 \cite{E6872} and ARGUS \cite{ARGUS2} collaborations.
In all these experiments the resonance appears as a pronounced peak 
in the invariant mass distribution of the $\pi^+\pi^-\Lambda_c^+$ channel.

It is now generally accepted that the $\Lambda_c$(2595) is the charmed
counterpart of the $\Lambda$(1405) \cite{PDG}. Therefore, it is natural
that interaction models of the $\bar KN$ system in which the $\Lambda$(1405)
appears as a dynamically generated state, as it is the case in chiral unitary
approaches as well as in the traditional meson-exchange picture, 
likewise generate the $\Lambda_c$(2595) dynamically, provided that SU(4) 
symmetry is assumed when extending the interactions from the strangeness  
to the charm sector. 
In this context it is important to realize that there are also drastic
differences between the strangeness and the charm case. 
In particular, 
the $\Lambda$(1405) is located fairly close to the $\bar KN$ threshold,
which is around 1430~MeV, while the $\Lambda_c$(2595) coincides practically
with the $\pi\Sigma_c$ threshold, which is at around 2593~MeV.
Furthermore, the $\pi\Sigma$ and $\bar KN$ thresholds are roughly 100~MeV apart,
while there are almost 200~MeV between the $\pi\Sigma_c$ and $DN$ thresholds. 
Finally, the $\pi\pi\Lambda_c$ channel -- where the $\Lambda_c$(2595) is 
experimentally observed -- opens 35~MeV below the resonance, while the corresponding 
$\pi\pi\Lambda$ channel is barely open at the location of the $\Lambda$(1405). 
Note that the $\pi\pi$ channel must be in the $I^G(J^P) = 0^+(0^+)$ (``$\sigma$'')
state when $\pi\pi\Lambda_c$ couples to the $\Lambda_c$(2595). And, due to parity
conservation, it is the $P_{01}$ partial wave of the 
${\pi\pi}\Lambda_c$ (${\pi\pi}\Lambda$) system which couples to the 
$S_{01}$ $DN$ ($\bar KN$) channel. 

\begin{figure*}[t]
\begin{center}
\includegraphics[height=100mm]{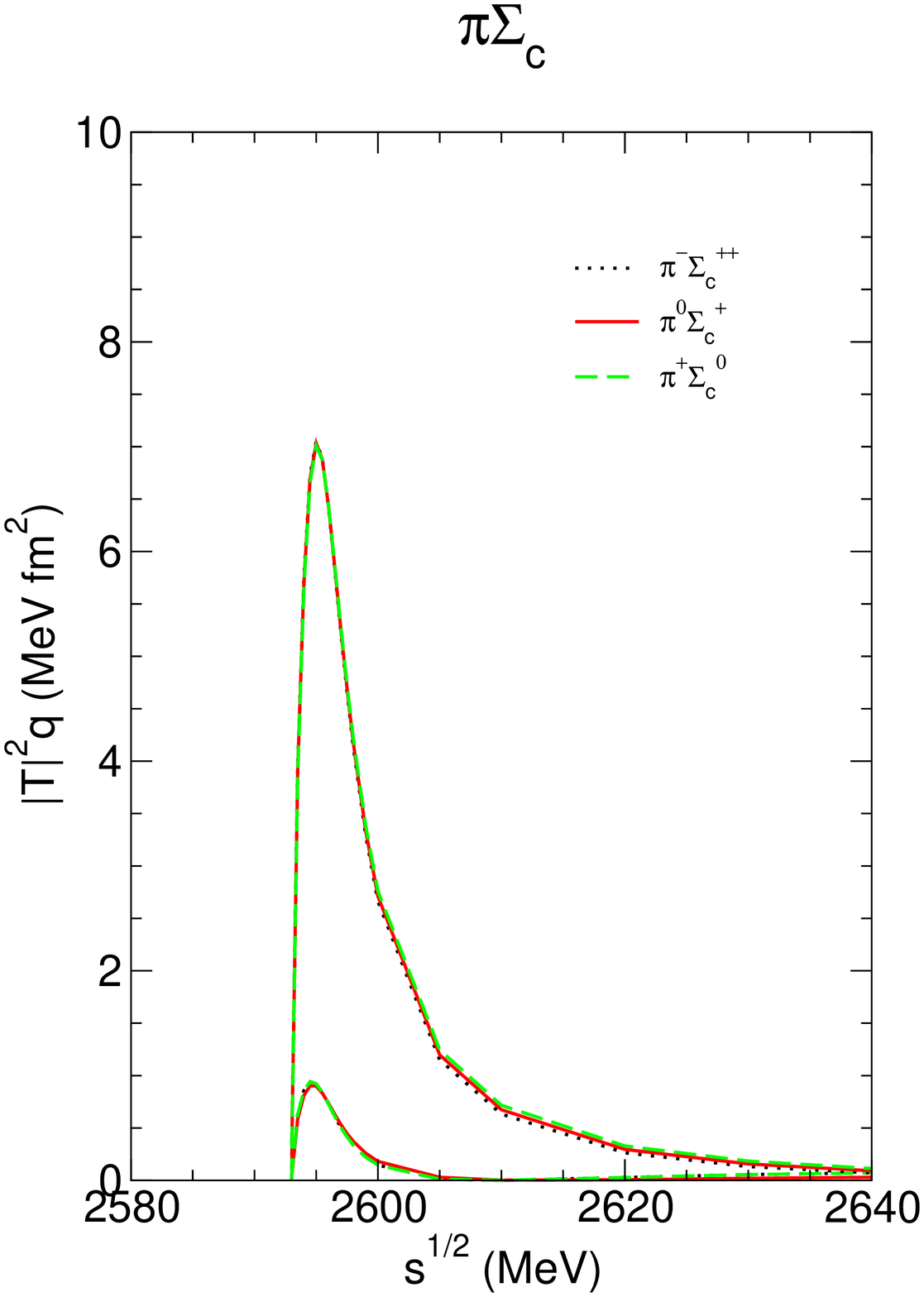}
\includegraphics[height=100mm]{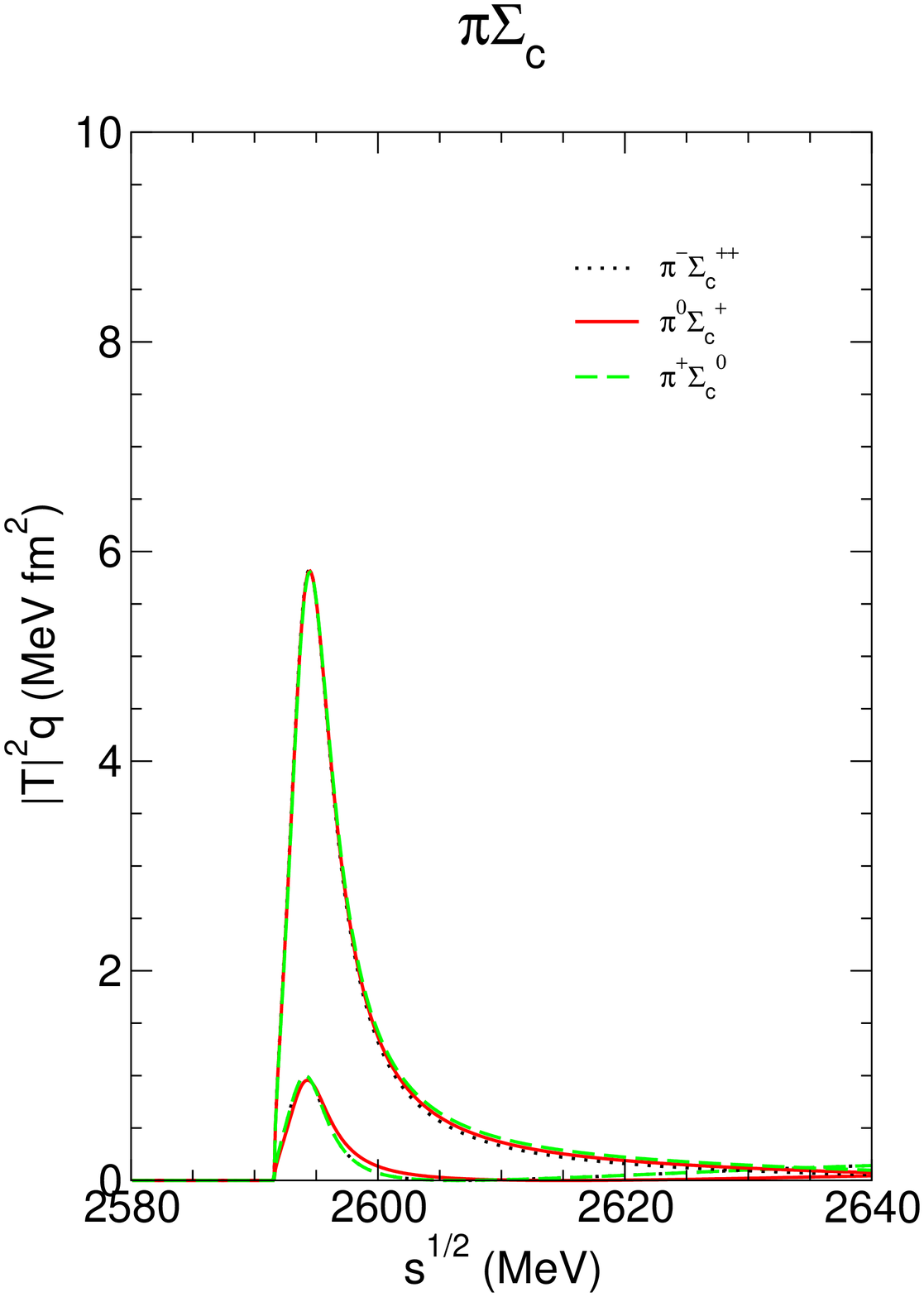}
\caption{$\pi\Sigma_c$ invariant mass spectrum in an 
isospin-symmetric calculation. Left are results based on our meson-exchange 
model while those on the right are for the SU(4) WT model of Ref.~\cite{Mizutani06}.
The lower curves are based on the $\pi\Sigma_c \to \pi\Sigma_c$
$T$-matrix while the upper ones correspond to $D^0p\to \pi\Sigma_c$.
}
\label{fig:3}
\end{center}
\end{figure*}

In view of the mentioned kinematical differences, it is certainly not surprising 
that the models do not really predict the $\Lambda_c$(2595) at exactly the 
position where it was found in the experiment. A fine-tuning
of inherent parameters such as subtraction constants or coupling constants
is required to shift the resonance to the observed energy. 
Specifically, in case of the J\"ulich $DN$ model the coupling constants of the 
scalar mesons were adjusted in such a way that the $\pi\Sigma_c$ $S$-wave phase 
shift in the $I=0$ channel goes through 90$^\circ$ at 2595~MeV, i.e. at the 
nominal $\Lambda_c$(2595) mass as listed in the PDG \cite{PDG}. 
It should be said, however, that for an investigation of the $DN$ interaction, 
as performed in Ref.~\cite{Haiden08}, the precise position of the 
$\Lambda_c$(2595) resonance does not play a role.  

The experimental papers report uniformly that the $\Lambda_c$(2595) 
decays dominantly into the $\pi^+\Sigma_c^0$ and $\pi^-\Sigma_c^{++}$ 
channels \cite{E6872,ARGUS2,CLEO2}. In the latter reference one can even
read that $\Lambda_c$(2595) decays almost 100\% to $\pi\Sigma_c$. 
At first sight this seems not unreasonable. Considering the reported mass 
difference $M(\Lambda_c(2595)) - M(\Lambda_c)$ of 
\begin{eqnarray}
307.5\pm 0.5 \pm 1.2 \ &{\rm MeV  \ [CLEO \ '95]}, \nonumber \cr
309.7\pm 0.9 \pm 0.4 \ &{\rm MeV  \ [E687 \ '96]}, \nonumber \cr
309.2\pm 0.7 \pm 0.3 \ &{\rm MeV  \ [ARGUS \ '97]}, 
\end{eqnarray}
respectively, and the corresponding threshold values for the $\pi\Sigma_c$ 
channels 
\begin{eqnarray}
M(\pi^-)+M(\Sigma_c^{++})- M(\Lambda_c) &= 307.13\pm 0.18 \ {\rm MeV}, \nonumber \cr
M(\pi^0)+M(\Sigma_c^{+})- M(\Lambda_c) &= 301.42\pm 0.4 \ {\rm MeV}, \nonumber \cr
M(\pi^+)+M(\Sigma_c^{0})- M(\Lambda_c) &= 306.87\pm 0.18 \ {\rm MeV}, 
\end{eqnarray}
where we use the latest values from the PDG \cite{PDG}, there is some 
phase space for the $\Lambda_c(2595) \to \pi\Sigma_c$ decay. 

However, the new CLEO measurement with improved statistics and with better
momentum resolution \cite{CLEO2} suggests a mass difference of only 
\begin{eqnarray}
305.3 \pm 0.4 \pm 0.6 \ &{\rm MeV  \ [CLEO \ '99]}.
\end{eqnarray}
A very similar mass difference
(305.6$\pm$0.3 MeV) was obtained in an independent analysis of the new
CLEO data by Blechman et al. \cite{Blechman}. Such a value reduces 
the phase space for the decay of the $\Lambda_c(2595)$ into the 
$\pi^+\Sigma_c^0$ and $\pi^-\Sigma_c^{++}$ channels significantly. 
Indeed the decay is only possible due to the finite widths of the 
involved particles. But, since the widths are only in the order of 
2~MeV ($\Sigma_c^{++}$, $\Sigma_c^{0}$) to 4~MeV $\Lambda_c(2595)$
\cite{PDG} and the detector resolution is 1.28~MeV \cite{CLEO2},
it is still surprising that the $\Lambda_c(2595)$ should decay 
domaninatly into those two channels as suggested by the experiment. 
We will come back to this issue at the end of this section. 

In the following we present predictions of the J\"ulich $DN$ model
and of the SU(4) WT model of Ref.~\cite{Mizutani06} for the invariant
mass distributions, i.e. for the quantity $|T|^2\cdot\, q$, 
in the relevant $\pi\Sigma_c$ channels. 
This allows us to explore whether there are any quantitative or even 
qualitative differences in the predictions of those models. 
Furthermore, we can illustrate the subtle effects of the slightly 
different thresholds of the 
$\pi^+\Sigma_c^0$, $\pi^0\Sigma_c^+$, and $\pi^-\Sigma_c^{++}$ 
channels on the various invariant mass distributions due to
the presense of a near-by pole. 
But first, let us show results for the isospin-symmetric calculation 
(based on averaged masses), cf. Fig.~\ref{fig:3}.
The upper curves correspond to the $D N \to \pi\Sigma_c$ $T$-matrix
while the lower ones are for $\pi\Sigma_c \to \pi\Sigma_c$.
In this figure 
the relative normalization of the $\pi\Sigma_c$ to the $DN\to \pi\Sigma_c$ 
channel is kept as predicted by the models but all $T$-matrices are
multiplied with the mass factor $M_\pi M_{\Sigma_c} / (M_\pi + M_{\Sigma_c})$ 
in order to obtain convenient units for plotting. 
 
\begin{figure*}[t]
\begin{center}
\includegraphics[height=100mm]{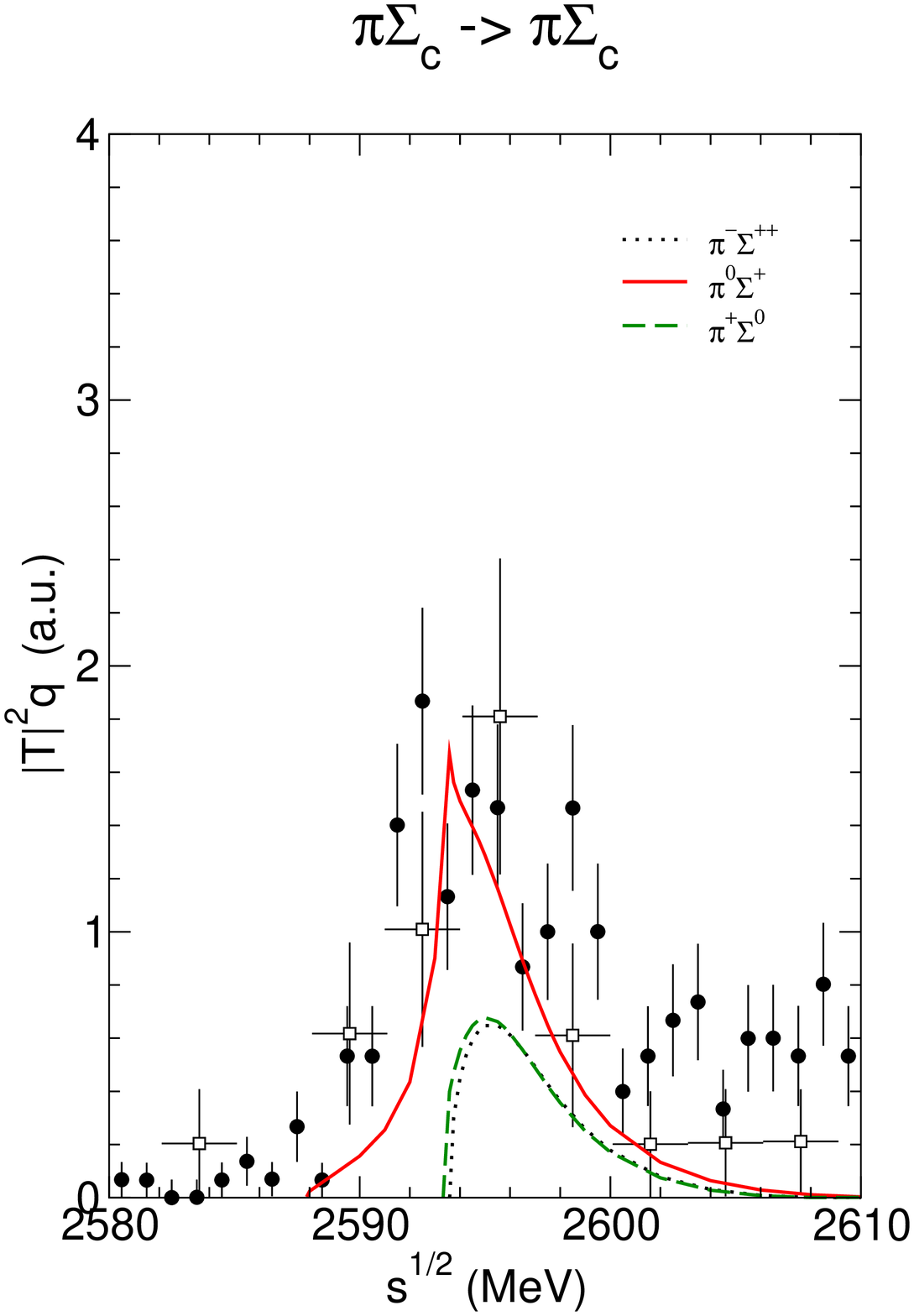}
\includegraphics[height=100mm]{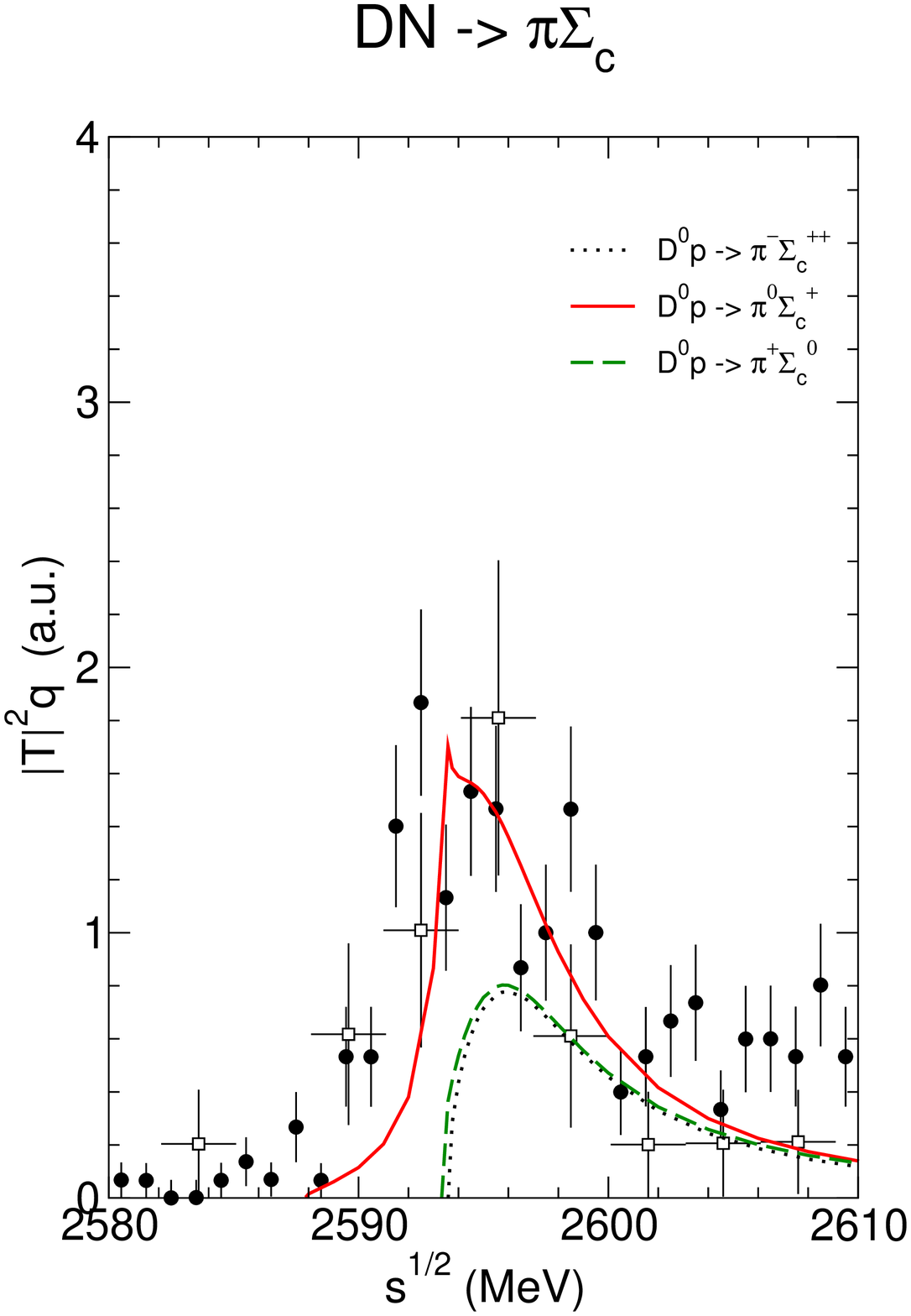}
\caption{$\pi\Sigma_c$ invariant mass spectrum predicted by our 
$DN$ meson-exchange model.
Left are results based on the $\pi\Sigma_c \to \pi\Sigma_c$ 
$T$-matrix and right the ones for $D^0p\to \pi\Sigma_c$. 
For illustration purposes we show also data for the 
$\pi^+\pi^-\Lambda^+_c$ invariant mass distribution 
taken from \cite{ARGUS2} (squares) and \cite{CLEO2} (circles).
}
\label{fig:8}
\end{center}
\end{figure*}

The results shown in Fig.~\ref{fig:3} make clear 
that the amplitudes of the $D N$-$\pi\Sigma_c$ systems are 
completely dominated by the $I=0$ 
contribution in the region of the $\Lambda_c(2595)$. 
As a consequence, the predictions for all charge states practically 
coincide. As mentioned above, this is not the case for $\bar K N$-$\pi\Sigma$
where the interference between the $I=0$ and $I=1$ amplitudes is significant.
For example, there the invariant mass distribution for 
$\pi^\pm \Sigma^\mp \to \pi^\pm\Sigma^\mp$ differs by roughly a factor of
two from that of $\pi^0 \Sigma^0 \to \pi^0\Sigma^0$, cf. Fig.~\ref{fig:2}. 
Furthermore, it can be seen from Fig.~\ref{fig:3} that the invariant mass 
distribution obtained from the $\pi\Sigma_c$ and $DN\to \pi\Sigma_c$ 
$T$-matrices are fairly similar. 
In fact, a comparison of the results shown in the left and right 
panels of Fig.~\ref{fig:3} reveals that there is even hardly any
difference between the results of the meson-exchange model and the
WT interaction. Even the absolute magnitudes are rather similar. 
This might be surprising but is certainly a reflection of the 
specific situation with the $\Lambda_c(2595)$ being located very
close to the $\pi\Sigma_c$ threshold. In such cases one expects 
to see features that are practically model-independent 
\cite{Hanhart10,Baru04,Baru05}.
In this context let us mention that the $DN\to \pi\Sigma_c$ results 
of the J\"ulich model are calculated from the half-off-shell
transition $T$-matrix assuming the $DN$ momentum to be zero, while for
the WT result \cite{Mizutani06} the on-shell amplitude is used. In the
latter case the corresponding $DN$ momentum is purely imaginary. 

Results based on the physical masses are presented in Fig.~\ref{fig:8}.
Here the invariant mass distributions are shown in arbitrary units and 
normalized in such
a way that one can easily compare the results based on the 
$\pi\Sigma_c\to \pi\Sigma_c$ (left panel) and $DN\to \pi\Sigma_c$ 
(right panel) $T$-matrices. But we keep the relative normalization 
between the different charge channels as predicted by the model. 
Obviously, there are drastic effects due to the different thresholds. 
The threshold of the $\pi^0\Sigma_c^+$ channel is about 6~MeV
lower than those of the other two charge channels and, as a consequence, the
predicted invariant mass distribution is about twice as large.
Moreover, there is a clear cusp in $\pi^0 \Sigma_c^+$ 
at the opening of the $\pi^+\Sigma_c^0$ channel. The threshold of 
$\pi^-\Sigma_c^{++}$ is just about 0.3 MeV above the one for $\pi^+\Sigma_c^0$.
It produces a noticeable kink in the $\pi^0 \Sigma_c^+$ 
invariant mass distribution. 
On the other hand, and as already anticipated from the comparison of the
isospin-symmetry calculation above, there are only rather subtle differences
between the lines shapes predicted from the $\pi\Sigma_c\to \pi\Sigma_c$ and 
from the corresponding $DN\to \pi\Sigma_c$ amplitudes. The only more
qualitative difference consists in the stronger fall-off with increasing
invariant mass exhibited by the results based on the $\pi\Sigma_c\to \pi\Sigma_c$
$T$-matrix. 
 
\begin{figure*}
\begin{center}
\includegraphics[height=100mm]{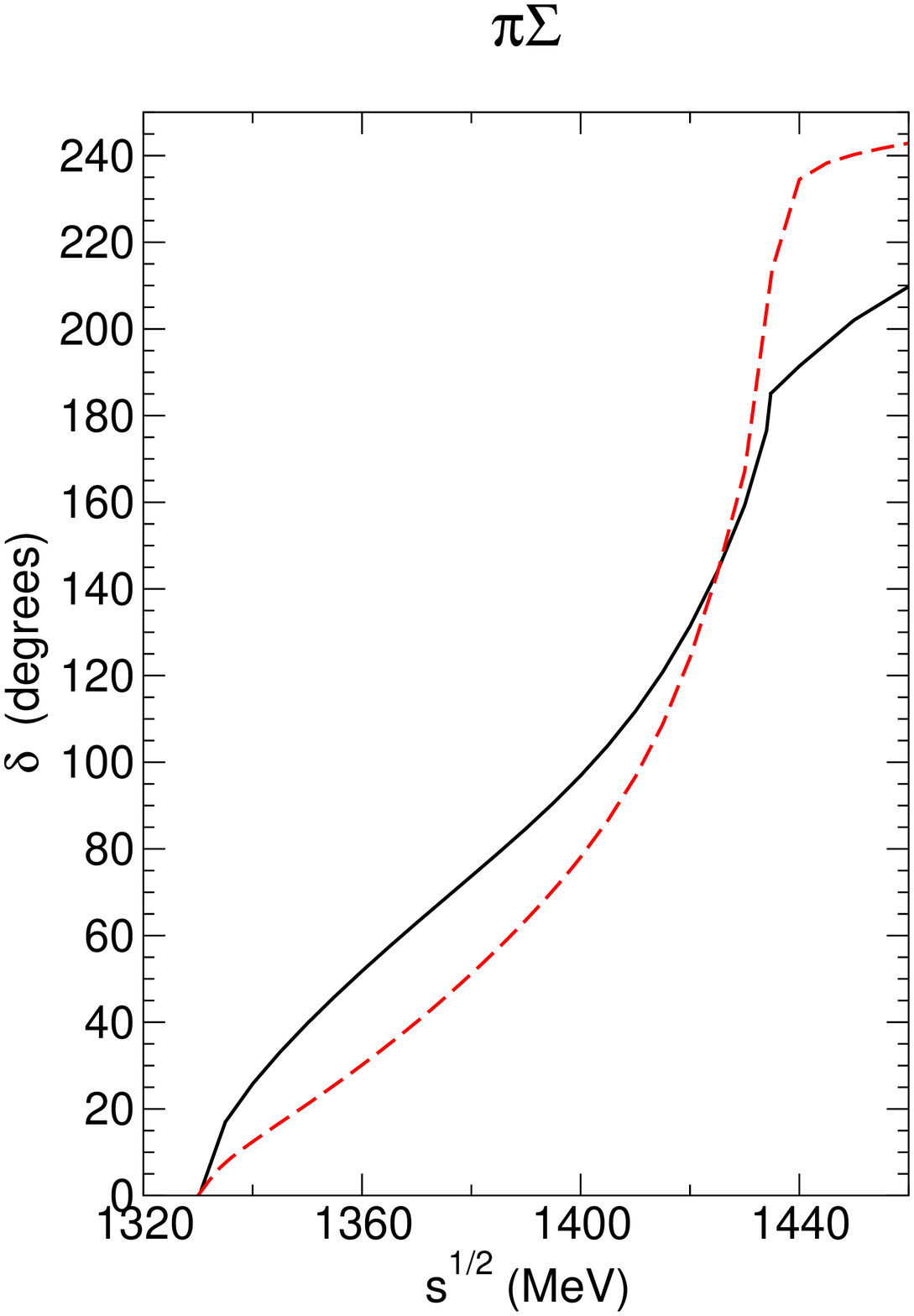}
\includegraphics[height=100mm]{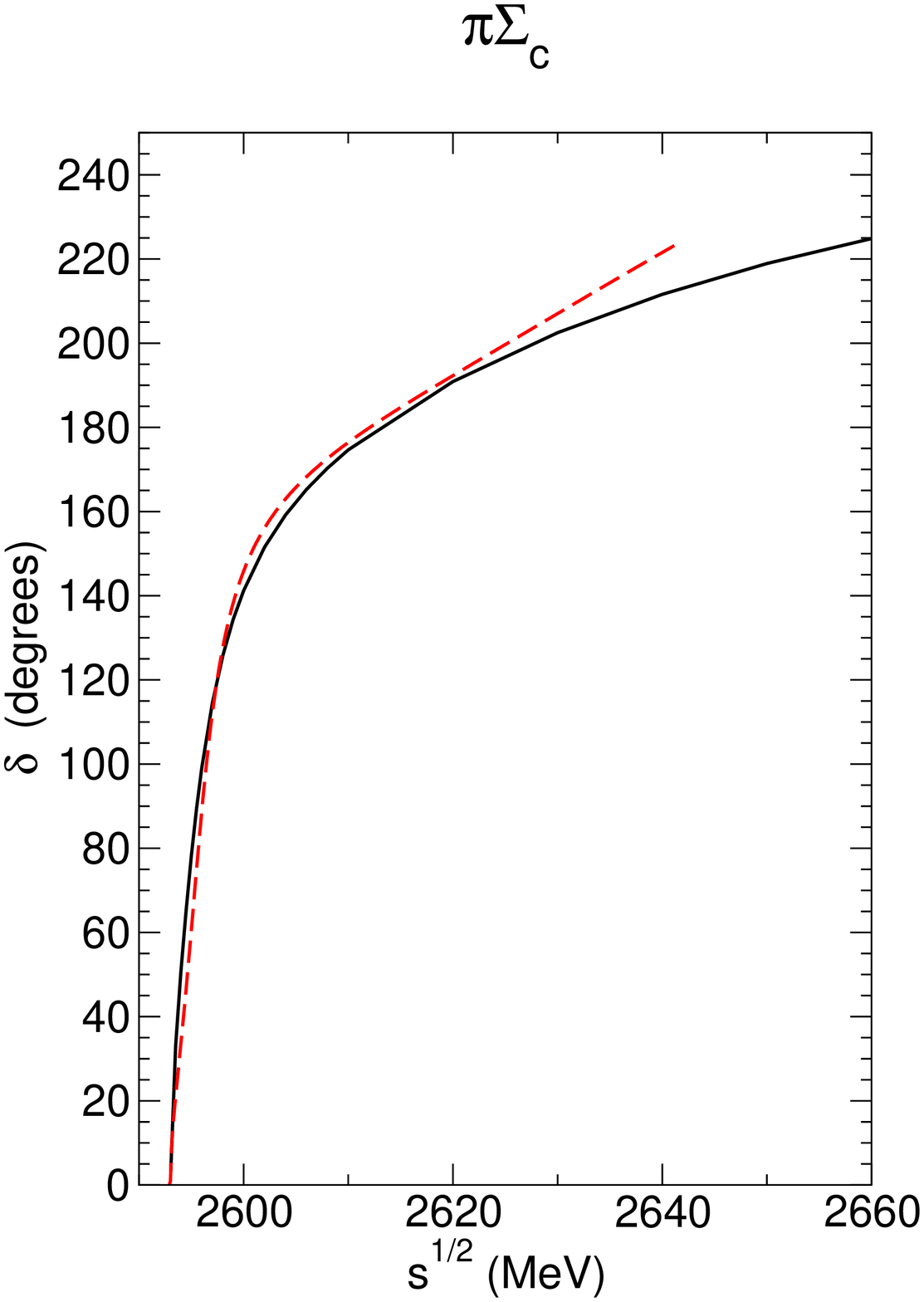}
\caption{$S_{01}$ $\pi\Sigma$ and $\pi\Sigma_c$ phase shifts.
Solid lines represent results for the J\"ulich $\bar KN$ and 
$DN$ models while the dashed lines are those for the corresponding
WT interactions \cite{Mizutani06,OR98}.
}
\label{fig:1a}
\end{center}
\end{figure*}

Finally, let us come to the experimental $\pi^+\pi^-\Lambda^+_c$
mass distribution where the signal for the $\Lambda$(2595)
was found. The corresponding data \cite{ARGUS2,CLEO2} are 
also displayed in Fig.~\ref{fig:8}. 
The results for the $D^0p \to \pi^0\Sigma_c^+$ as well as 
for the $\pi^0\Sigma_c^+ \to \pi^0\Sigma_c^+$ channels resemble
indeed very much the measured signal 
and one can imagine that smearing out our results by the width of
the $\Sigma_c^+$, which is roughly 4~MeV \cite{CLEO2,PDG}, would 
yield a fairly good fit to the data. However, experimentally it was
found that the $\Lambda_c$(2595) decays predominantly into the
$\pi^+\Sigma_c^0$ and $\pi^-\Sigma_c^{++}$ channels with a branching 
fraction in the range of 66\% \cite{ARGUS2} to close to 100\% \cite{CLEO2}.
Smearing out the corresponding results with the significantly
smaller and better known widths of the $\Sigma_c^0$ and $\Sigma_c^{++}$,
of just 2~MeV \cite{PDG}, would still leave many of the events 
found below the nominal $\pi^+\Sigma_c^0$ and $\pi^-\Sigma_c^{++}$ 
threshold unexplained, especially for the CLEO experiment \cite{CLEO2}.  
Thus, it seems to us that the position of the $\Lambda_c(2595)$ resonance
being so close to or even below the nominal $\pi^+\Sigma_c^0$/$\pi^-\Sigma_c^{++}$
thresholds and the found large branching ratios into those channels
are difficult to reconcile. 

Independently of that, we would like to say also a word of caution 
concerning our own results. In view of the fact that the signal for the 
$\Lambda_c(2595)$ resonance is seen in the $\pi^+\pi^-\Lambda_c^+$ 
channel, any more rigorous model analysis would definitely require 
the explicit inclusion of this channel. 
In principle, the presence of the $\pi\pi \Lambda_c^+$ channel 
could be simulated within our model by adding a phenomenological 
$\sigma\Lambda_c^+$ channel, analogous to the treatment of the 
$\pi\pi N$ channel in our $\pi N$ model \cite{Krehl,Achot}.
But then many new parameters would have to be introduced that can no 
longer be fixed by SU(4) arguments in a reasonable way. 

In any case, first it would be important to confirm the new 
CLEO data by independent measurements of the $\pi\pi\Lambda_c^+$
and $\pi\Sigma_c$ mass spectra in the region of the $\Lambda_c(2595)$. 
Specifically, it would be essential to establish unambigously that 
there is a large decay rate of that resonance into the $\pi\Sigma_c$
channels. 
A precise determination of the pole position of the $\Lambda_c(2595)$
could then be done along model-independent approaches such as the
ones suggested in Refs.~\cite{Blechman,Hanhart10}. Besides of being much
better suited for performing a fit to data and for deducing uncertainties, 
these approaches allow one to incorporate also finite widths effects 
appropriately which is very difficult to achieve in models like the 
ones discussed in the present paper. In view of the experimental 
situation discussed above such finite widths effects might play a 
crucial role. 

Note that the predictions of the SU(4) WT model for the 
quantities $|T_{\pi\Sigma_c \to \pi\Sigma_c}|^2\cdot q$ 
and $|T_{D^0 p \to \pi\Sigma_c}|^2\cdot q$ are very similar to the
ones of the meson-exchange model and, therefore, we do not show the 
corresponding curves here. 

In order to understand the differences in the mass spectra
for the strangeness and charm sectors it is instructive to take a 
look at the phase shifts of the $\pi\Sigma$ and $\pi\Sigma_c$ channels 
in the $S_{01}$ partial wave where the poles corresponding to the 
$\Lambda(1405)$ and $\Lambda_c(2595)$ are located. 
Corresponding results are presented in Fig.~\ref{fig:1a}.
The standard relation of the (partial-wave projected) $T$-matrix 
to the phase shift is $T(q) = - \exp(i\delta(q)) \sin (\delta(q))/q$. 
The quantity $\sin^2 (\delta(q))$ has its maximum where 
$\delta(q)$ passes through 90$^\circ$. The maximum of the corresponding
invariant mass distribution, $|T|^2\cdot q=\sin^2 (\delta(q))/q$, 
will occur at somewhat smaller invariant masses, 
due to the additional $1/q$ factor and
depending on the slope with which the phase shift passes through 
90$^\circ$. The invariant mass distribution will be zero where the phase 
shift passes through 180$^\circ$. 
These are indeed the features of the $\pi\Sigma \to \pi\Sigma$ invariant 
mass spectrum, shown in Fig.~\ref{fig:2} on the left side. 
On the other hand, the $\bar K N \to \pi\Sigma$ invariant mass spectrum 
has its maximum close to the position of the pole. This corresponds
to the region where the $\pi\Sigma$ phase shift exhibits
the strongest variation with energy, which is around
1420 to 1430~MeV, cf. Fig.~\ref{fig:1a} (left). 
 
In this context we would like to draw attention to the fact
that the behavior of the $\pi\Sigma$ $S_{01}$ phase shift and
the pertinent invariant mass distribution is very similar to
the one of the $\pi\pi$ $\delta_{00}$ partial wave. 
In the latter case the phase shift shows a broad shoulder at lower 
energies, passing slowly
through 90$^\circ$, a behavior usually associated with the $\sigma$
meson (or $f_0(600)$), while finally rising quicky through 180$^\circ$ 
around 1~GeV at the location of the $f_0(980)$ resonance.
The corresponding mass spectrum 
consists in a broad bump on which the $f_0(980)$ appears as a
dip structure, cf. Ref.~\cite{Gunter01}. Also in case of the 
$\bar K N$-$\pi\Sigma$ system the pole at around 1436~MeV with the
smaller width, which one might associate with the $\Lambda$(1405),
produces a peak in the $\bar K N\to \pi\Sigma$ invariant mass spectrum
but a dip in the one computed from the $\pi\Sigma\to\pi\Sigma$  
$T$-matrix. 

The behavior of the corresponding phase shift for the
$\pi\Sigma_c$ system is very much different, cf. Fig.~\ref{fig:1a} (right).
Here the strongest variation with energy occurs already very close to the 
threshold and in the same region the phase also goes through 90$^\circ$. 

The phase shift predicted by the SU(4) WT model \cite{Mizutani06,OR98} is shown 
by the dashed line in Fig.~\ref{fig:1a}. Obviously, for the 
$DN$-$\pi\Sigma_c$ system it is very similar to the prediction of the 
meson-exchange model. Thus, it is not surprising that also the corresponding 
invariant mass distributions, presented in Fig.~\ref{fig:8}, are very similar 
to those of the J\"ulich potential. 
For $\bar KN$-$\pi\Sigma$ there are noticeable quantitative differences.
In particular, the slope of the phase shift is significantly larger where 
it passes through 90$^\circ$, reflecting the fact that the two poles of the WT
model \cite{OR98} are much closer to each other \cite{Jido:2003cb} than 
in case of the J\"ulich model, and, as a consequence, the maxima of the
invariant mass spectra based on the $\bar KN \to \pi\Sigma$ and 
$\pi\Sigma \to \pi\Sigma$ amplitudes are also closer to each 
other \cite{Jido:2003cb}. 

\section{Summary}

In this paper we presented a model for the interaction in the 
coupled systems $DN$, $\pi\Lambda_c$, and $\pi\Sigma_c$,
which was developed in close analogy to
the meson-exchange $\bar KN$ interaction of the J\"ulich group \cite{MG},
utilizing SU(4) symmetry constraints. The main ingredients of
the $DN$ interaction are provided by vector meson ($\rho$, $\omega$)
exchange but higher-order box diagrams involving ${D}^*N$,
$D\Delta$, and ${D}^*\Delta$ intermediate states, are also included.
The coupling of the $DN$ system to the $\pi\Lambda_c$ and 
$\pi\Sigma_c$ channels is facilitated by $D^*$(2010) exchange
and by nucleon $u$-channel pole diagrams. 

The interaction model generates several states dynamically. 
The narrow $DN$ quasi-bound state found in the $S_{01}$
partial wave is identified with the ($I=0$) $\Lambda_c$(2595) 
resonance.
Narrow states were also found in the $S_{11}$ and $P_{01}$ partial 
waves. We identify the former with the $I=1$ resonance $\Sigma_c(2800)$ 
and the latter with the $\Lambda_c(2765)$ resonance, whose quantum
numbers are not yet established~\cite{PDG}.
 
Results for $DN$ total and differential cross sections were presented
and compared with predictions of an interaction model that
is based on the leading-order Weinberg-Tomozawa term 
\cite{Mizutani06}. 
While the predictions of the two models for the $I=1$ channel 
are fairly similar, in magnitude as well as with regard to the
energy dependence, this is not the case for $I=0$ amplitude. Here
the possible presence of a $P$-wave resonance near the
threshold, i.e. the $\Lambda_c(2765)$, has a dramatic influence
on the shape and the energy dependence of the cross section. 
 
Finally, we discussed the $\Lambda_c$(2595) resonance and 
the role of the near-by $\pi\Sigma_c$ threshold. In particular,
we presented results for the $\pi\Sigma_c$ invariant mass spectrum
in the particle basis which illustrate the subtle effects of the 
slightly different thresholds of the $\pi^+\Sigma_c^0$, $\pi^0\Sigma_c^+$, 
and $\pi^-\Sigma_c^{++}$ channels on the various invariant mass 
distributions. 
We also pointed out that there seems to be a contradiction between the
observation that the narrow $\Lambda_c$(2595) resonance decays
almost exclusively into the $\pi^+\Sigma_c^0$ and $\pi^-\Sigma_c^{++}$ 
channels and the latest values of its mass, which place the resonance 
about 2~MeV below the thresholds of those channels \cite{CLEO2,Blechman}. 
Indeed with a mass of 2592.06 $\pm$ 0.3 MeV, as determined in 
Ref.~\cite{Blechman}, the $\Lambda_c$(2595) would lie just between 
the $\pi^0\Sigma_c^+$ and the $\pi^+\Sigma_c^0$/$\pi^-\Sigma_c^{++}$
thresholds, which surely would be an interesting scenario. 

\vskip 0.2cm
\noindent
{\bf Acknowledgments}

\smallskip\noindent
J.H. acknowledges instructive discussions with
Michael D\"oring about the determination of poles in the complex
plane.  This work is partially supported by the Helmholtz Association through funds
provided to the virtual institute ``Spin and strong QCD'' (VH-VI-231), by
the EU Integrated Infrastructure Initiative  HadronPhysics2 Project (WP4
QCDnet), by BMBF (06BN9006) and by DFG (SFB/TR 16, ``Subnuclear Structure of Matter''). 
G.K. acknowledges
financial support from CAPES, CNPq and FAPESP (Brazilian agencies). 
L.T. wishes to acknowledge support from the Rosalind Franklin Programme 
of the University of Groningen (The Netherlands) and the Helmholtz
International Center for FAIR within the framework of the LOEWE
program by the State of Hesse (Germany).

\appendix

\section{The interaction Lagrangians}
\label{app:lags}

In this appendix we list the specific interaction Lagrangians which are used to
derived the meson-exchange $\bar DN$ interaction. The
baryon-baryon-meson couplings are given by \cite{MG} 
\begin{eqnarray}
\nonumber
{\cal L}_{BBS} &=& g_{BBS} \bar \Psi_B(x) \Psi_B(x) \Phi_S(x) \ , \\
\nonumber
{\cal L}_{BBP} &=& g_{BBP} \bar \Psi_B(x) i\gamma^5 \Psi_B(x) \Phi_P(x) \ , \\
\nonumber
{\cal L}_{BBV} &=& g_{BBV} \bar \Psi_B(x) \gamma_\mu \Psi_B(x) \Phi_V^\mu (x) \\
\nonumber
&+& \frac{f_{BBV}}{4 m_N} \bar \Psi_B(x) \sigma_{\mu\nu} \Psi_B(x)
(\partial^\mu \Phi_V^\nu (x) - \partial^\nu \Phi_V^\mu (x)) \ , \\
\nonumber
{\cal L}_{B\Delta P} &=& \frac{f_{B\Delta P}}{m_P} \bar \Psi_{\Delta \mu}(x) \Psi_B(x)
\partial^\mu \Phi_P (x) + H.c. \ , \\
\nonumber
{\cal L}_{B\Delta V} &=& \frac{f_{B\Delta V}}{m_V} i (\bar \Psi_{\Delta \mu}(x) \gamma^5 \gamma_\mu \Psi_B(x) \\
&-& \bar \Psi_B(x) \gamma^5 \gamma_\mu \Psi_{\Delta \mu}(x))
(\partial^\mu \Phi_V^\nu (x) - \partial^\nu \Phi_V^\mu (x)) \ . 
\end{eqnarray}
Here, $\Psi_B$ and $\Psi_{\Delta\mu}$ are the nucleon (or hyperon) and $\Delta$
field operators and $\Phi_S$, $\Phi_P$, and $\Phi_V^\mu$ are the field operators for scalar,
pseudoscalar and vector mesons, respectively.

\begin{table*}[ht]
\caption{Vertex parameters used in the meson-exchange model of the
$DN$ interaction.  $m_{exch}$ is the mass of the exchanged particle. 
$g_M$ and $g_B$/$f_B$ ($\Lambda_M$ and $\Lambda_B$) refer to the 
coupling constants (cutoff masses) used at the meson-meson-meson and 
baryon-baryon-meson 
(for pole and baryon-exchange graphs: upper and lower) vertex,
respectively. 
The $\sigma$ and $a_0$ coupling constants in brackets are those 
that follow from our $\bar DN$ model \cite{Haiden07} under the
assumption of SU(4) symmetry. 
}
 \begin{center}
\renewcommand{\arraystretch}{1.2}
\begin{tabular}{|c|cccccc|}
 \hline
Process & Exch. part. & $m_{exch}$
 & $g_M g_B/(4 \pi)$ & $g_M f_B /(4 \pi)$ 
 & $\Lambda_M$  & $\Lambda_B$   \\
        &  & [MeV]   &  & & \ [GeV] \ & \ [GeV] \ \\
\hline $ D N \rightarrow  D N$
                     & $\rho$         & \phantom{1}769\phantom{.03}
 &\phantom{--4}0.773 &\phantom{--4}4.713 
 & 1.4 & 1.6  \\
                     & $\omega$       & \phantom{1}782.6\phantom{3}
 &\phantom{--4}-2.318  &\phantom{--4}0.0
 & 1.5 & 1.5  \\
 & $\sigma$       & \phantom{1}600\phantom{.03}
 &\phantom{--4}2.60 [1.00] &\phantom{--4}-
 & 1.7 & 1.2  \\
 & $a_0$       & \phantom{1}980\phantom{.03}
 &\phantom{--4}-4.80 [-2.60] &\phantom{--4}-
 & 1.5 & 1.5  \\
                     & $\Lambda_c$      & 2286.5
 &\phantom{-4}15.55 &\phantom{--4}- 
 & 1.4 & 1.4  \\
                     & $\Sigma_c$       & 2455\phantom{.03}
 &\phantom{--4}0.576 &\phantom{--4}-
 & 1.4 & 1.4  \\
$D N \rightarrow {D}^* N$ & $\pi$          &
\phantom{1}138.03
 &\phantom{--4}3.197 &\phantom{--4}-
 & 1.3 & 0.8  \\
                        & $\rho$         & \phantom{1}769\phantom{.03}
 &\phantom{--4}-0.773  &\phantom{--4}-4.713 
 & 1.4 & 1.0  \\
$D N \rightarrow {D}^* \Delta$ & $\pi$          &
\phantom{1}138.03
 &\phantom{--4}0.506 &\phantom{--4}-
 & 1.2 & 0.8  \\
                          & $\rho$         & \phantom{1}769\phantom{.03}
 &\phantom{--4}-4.839 &\phantom{--4}- 
 & 1.3 & 1.0  \\
$D N \rightarrow D  \Delta$ & $\rho$         &
\phantom{1}769\phantom{.03}
 &\phantom{--4}4.839 &\phantom{--4}- 
 & 1.3 & 1.6 \\
 \hline
$D N \rightarrow \pi \Lambda_c$ & $D^*$         &
2009\phantom{.03}
 &\phantom{--4}-1.339 &\phantom{--4}-4.365 
 & 3.1 & 3.1 \\
                     & $N$      & 938.926
 &\phantom{-4}-14.967 &\phantom{-4}- 
 & 2.5 & 1.4  \\
                     & $\Sigma_c$       & 2455\phantom{.03}
 &\phantom{--4}1.995 &\phantom{--4}- 
 & 3.5 & 1.4  \\
 \hline
$D N \rightarrow \pi \Sigma_c$ & $D^*$         &
2009\phantom{.03}
 &\phantom{--4}-0.773 &\phantom{--4}1.871
 & 3.1 & 3.1 \\
                     & $N$      & 938.926
 &\phantom{-4}2.88 &\phantom{-4}- 
 & 2.5 & 1.4  \\
                     & $\Lambda_c$      & 2286.5
 &\phantom{-4}-10.368 &\phantom{-4}- 
 & 2.8 & 1.4  \\
                     & $\Sigma_c$       & 2455\phantom{.03}
 &\phantom{--4}2.304 &\phantom{--4}- 
 & 3.5 & 1.4  \\
 \hline
$\pi \Lambda_c \rightarrow \pi \Lambda_c$ 
                     & $\Sigma_c$       & 2455\phantom{.03}
 &\phantom{--4}6.912 &\phantom{--4}- 
 & 3.5 & 3.5  \\
 \hline
$\pi \Lambda_c \rightarrow \pi \Sigma_c$ & $\rho$         & 
\phantom{1}769\phantom{.03}
 &\phantom{--4}0.0   &\phantom{--4}7.605
 & 2.0 & 1.35 \\
                     & $\Sigma_c$       & 2455\phantom{.03}
 &\phantom{--4}7.891 &\phantom{--4}- 
 & 3.5 & 3.5  \\
 \hline
$\pi \Sigma_c \rightarrow \pi \Sigma_c$ & $\rho$         & 
\phantom{1}769\phantom{.03}
 &\phantom{--4}3.092 &\phantom{--4}5.689
 & 2.0 & 1.16 \\
                     & $\Lambda_c$      & 2286.5
 &\phantom{-4}6.912 &\phantom{-4}- 
 & 2.8 & 2.8  \\
                     & $\Sigma_c$       & 2455\phantom{.03}
 &\phantom{--4}9.216 &\phantom{--4}- 
 & 3.5 & 3.5  \\
 \hline
\end{tabular}
\renewcommand{\arraystretch}{1.0}
 \end{center}
\label{coup}
\end{table*}

The employed three-meson couplings are
\begin{eqnarray}
\nonumber
{\cal L}_{PPS} &=& g_{PPS} m_P \Phi_P(x) \Phi_P(x) \Phi_S(x) \ , \\
\nonumber
{\cal L}_{PPV} &=& g_{PPV} \Phi_P(x) \partial_{\mu} \Phi_P(x) \Phi_V^{\mu}(x) \ , \\
\nonumber
{\cal L}_{VVP} &=& \frac{g_{VVP}}{m_V}i\epsilon_{\mu \nu \tau \delta}
\partial^\mu \Phi_V^\nu (x) \partial^\tau \Phi_V^\delta (x) \Phi_P(x) \ , \\
&&
\end{eqnarray}
where $\epsilon_{\mu \nu \tau \delta}$ is the totally  antisymmetric tensor in
four dimensions with $\epsilon^{0123}=1$. 
Details on the derivation of the meson-baryon interaction potential from those 
Lagrangians can be found in Ref.~\cite{MG} together with explicit expressions
for those potentials. 
The SU(4) flavour structure that leads to the characteristic 
relations between the coupling constants is discussed in 
Sect.~II of \cite{Haiden07}. 
All vertices are supplemented with form factors in order to suppress the 
meson-exchange contributions for high-momen\-tum transfer and guarantee
convergence when solving the Lipp\-mann-Schwinger equation. 
For all ($t$-channel) exchange diagrams those vertex form factors are 
parameterized in a conventional monopole form
\cite{MG} 
\begin{eqnarray}
F_\alpha (q^2) = \left( \frac{\Lambda^2_\alpha-m_{exch}^2}
{\Lambda^2_\alpha+{\bf q}^2} \right)^{n_\alpha}, 
\end{eqnarray}
where ${\bf q}^2$ is the square of the three-momentum transfer. Here $n_\alpha=1$
is used for all vertices except for the $N\Delta\rho$ vertex where $n_\alpha=2$
\cite{MG,MHE}. 
In case of ($s$ and $u$ channel) pole contributions a slightly different form is 
used to avoid problems of convergence and singularity, viz.
\begin{eqnarray}
F_\beta (q^2) = \left( \frac{\Lambda^4_\beta+m_{exch}^4}
{\Lambda^4_\beta+(q^2)^2} \right), 
\end{eqnarray}
where $q^2$ is the square of the four-momentum transfer \cite{MG}. 
Note that both forms are normalized in such a way that $F\equiv 1$ when
the exchanged particle is on its mass shell. 
The values for the vertex parameters (coupling constants and cutoff masses) that 
are used in our meson-exchange model of the $DN$ interaction are summarized
for convenience in Table~\ref{coup}. 
The following averaged masses are used for evaluating the interaction potential
and when solving the Lipp\-mann-Schwin\-ger equation in isospin basis:
$M_N = 938.926$~MeV, $M_{\Lambda_c} = 2286.5$~MeV, $M_{\Sigma_c} = 2455.0$~MeV,
$M_\pi = 138.03$~MeV, $M_D = 1866.9$~MeV. For the calculation in the particle basis
we use the masses as given in the PDG listing \cite{PDG}.


\begin{thebibliography}{20}

\bibitem{Lehmann:2009dx}
  I.~Lehmann  [PANDA Collaboration],
  arXiv:0909.4237 [hep-ex].


\bibitem{Staszel:2010zz}
  P.~Staszel  [CBM Collaboration],
  Acta Phys.\ Polon.\  B {\bf 41}, 341 (2010).

\bibitem{FAIR} http://www.gsi.de/fair/.


\bibitem{Haiden07}
	J. Haidenbauer, G. Krein, U.-G. Mei{\ss}ner, and  A. Sibirtsev,
	Eur. Phys. J. A {\bf 33}, 107 (2007).

\bibitem{Haiden08}
	J. Haidenbauer, G. Krein, U.-G. Mei{\ss}ner, and  A. Sibirtsev,
	Eur. Phys. J. A {\bf 37}, 55 (2008).

\bibitem{Matsui:1986dk}
  T.~Matsui and H.~Satz,
  Phys.\ Lett.\  B {\bf 178}, 416 (1986).

\bibitem{Thews}
  R.~L.~Thews, M.~Schroedter, and J.~Rafelski,
  Phys.\ Rev.\  C {\bf 63}, 054905 (2001).

\bibitem{Tsushima:1998ru}
  K.~Tsushima, D.~H.~Lu, A.~W.~Thomas, K.~Saito, and R.~H.~Landau,
  Phys.\ Rev.\  C {\bf 59}, 2824 (1999).

\bibitem{GarciaRecio:2010vt}
C.~Garcia-Recio, J.~Nieves and L.~Tolos,
Phys.\ Lett.\ B {\bf 690}, 369 (2010). 

\bibitem{Brodsky:1989jd}
  S.~J.~Brodsky, I.~A.~Schmidt, and G.~F.~de Teramond,
  Phys. Rev. Lett.  \textbf{64}, 1011 (1990).

\bibitem{Luke92}
M.E. Luke, A.V. Manohar, and M.J. Savage, 
Phys.\ Lett.\ B {\bf 288}, 355 (1992).

\bibitem{Krein:2010vp}
  G.~Krein, A.~W.~Thomas, and K.~Tsushima,
  arXiv:1007.2220 [nucl-th].

\bibitem{Ko:2000jx}
  S.~H.~Lee and C.~M.~Ko,
  Phys.\ Rev.\  C {\bf 67}, 038202 (2003). 

\bibitem{MG}
	A.~M\"uller-Groeling, K. Holinde, and J. Speth,
	Nucl. Phys. A {\bf 513}, 557 (1990).

\bibitem{Juel1} 
	R. B{\"u}ttgen, K. Holinde, A. M{\"u}ller--Groeling,
	J. Speth, and P. Wyborny, Nucl. Phys. A {\bf 506}, 586 (1990).

\bibitem{Juel2} 
	M. Hoffmann, J.W. Durso, K. Holinde, B.C. Pearce,
	and J. Speth, Nucl. Phys. A {\bf 593}, 341 (1995).

\bibitem{CLEO1} 
K.W. Edwards et al., Phys. Rev. Lett. {\bf 74}, 3331 (1995).

\bibitem{E6872} 
P.L. Frabetti et al., Phys. Lett. B {\bf 365}, 461 (1996).

\bibitem{ARGUS2} 
H. Albrecht et al., Phys. Lett. B {\bf 402}, 207 (1997).

\bibitem{CLEO2} 
 Jiu Zheng, Ph.D. dissertation, University of Florida, 1999,
http://www.lns.cornell.edu/public/THESIS/1999/
THESIS99-4/JiuZheng.ps

%

\bibitem{PDG}
	C. Amsler et al. (Particle Data Group), Phys. Lett. B
        {\bf 667}, 1 (2008).

\bibitem{Lutz04}
	M.~F.~M.~Lutz and E.~E.~Kolomeitsev,
 	Nucl.\ Phys.\  A {\bf 730}, 110 (2004).

\bibitem{Hofmann05}
        J.~Hofmann and M.~F.~M.~Lutz,
        Nucl.\ Phys.\ A {\bf 763}, 90 (2005).

\bibitem{Mizutani06}
	T.~Mizutani and A.~Ramos,
	Phys.\ Rev.\  C {\bf 74}, 065201 (2006).

\bibitem{Lutz06}
        M.~F.~M.~Lutz and C.~L.~Korpa,
        Phys.\ Lett.\  B {\bf 633}, 43 (2006).

\bibitem{GarciaRecio09} C.~Garcia-Recio, V.K.~Magas, T.~Mizutani, J.~Nieves, 
        A.~Ramos, L.L.~Salcedo, and L.~Tolos, 
        Phys. \ Rev. \ D {\bf 79}, 054004 (2009).

\bibitem{Tolos04}
	L.~Tolos, J.~Schaffner-Bielich, and A.~Mishra,
	Phys.\ Rev.\ C {\bf 70}, 025203 (2004).

\bibitem{Tolos08}
	L.~Tolos, A.~Ramos, and T.~Mizutani,
	Phys.\ Rev.\ C {\bf 77}, 015207 (2008).

\bibitem{hypx} for recent overviews see:
        W. Weise, Nucl.\ Phys.\ A {\bf 835}, 51 (2010);
        D. Jido et al., Nucl.\ Phys.\ A {\bf 835}, 59 (2010);
        Y. Aikaishi, T. Yamazaki, M. Obu, and M. Wada, 
        Nucl.\ Phys.\ A {\bf 835}, 67 (2010);
        A. Gal, Chinese Physics C {\bf 34(9)} (2010) 1169 
        [arXiv:0912.2214 [nucl-th]]. 

\bibitem{HHK}
	D.~Hadjimichef, J.~Haidenbauer, and G.~Krein,
	Phys.\ Rev.\ C {\bf 66}, 055214 (2002).

\bibitem{Dong}
	Y. Dong, A. Faessler, T. Gutsche, and V.E. Lyubovitskij, 
	Phys.\ Rev.\ D {\bf 81}, 074011 (2010).

\bibitem{Bora}
        B. Borasoy, U.-G. Mei{\ss}ner, and R. Ni{\ss}ler, 
        Phys.\ Rev.\ C {\bf 74}, 055201 (2006). 

\bibitem{MRR}
        U.-G.~Mei{\ss}ner, U.~Raha, and A.~Rusetsky,
         Eur.\ Phys.\ J.\  C {\bf 35}, 349 (2004). 

\bibitem{dear} G.~Beer {\it et al.} (DEAR Collaboration), 
        Phys. \ Rev. \ Lett.\ {\bf 94}, 212302 (2005); 
        M. Cargnelli {\it et al.} (DEAR Collaboration), 
        Int. \ J. \ Mod. \ Phys. \ A {\bf 20}, 341 (2005).

\bibitem{kek} M.~Iwasaki {\it et al.}, 
        Phys. \ Rev. \ Lett. \ {\bf 78}, 3067 (1997); 
        T.M.~ Ito {\it et al.}, Phys. \ Rev.\ C \ {\bf 58}, 2366 (1998).

\bibitem{siddharta} M.~Cargnelli {\it et al.}, 
        Nucl.\ Phys.\ A {\bf 835}, 27 (2010); M.~Cargnelli, 
        in {\it Proceedings of the 12th International Conference on 
        Meson-Nucleon Physics and the Structure of the Nucleon
        May 31-June 4, 2010, MENU2010}, AIP Conf. Proc., in preparation. 

\bibitem{Nowak} 
        R.J. Nowak et al., Nucl. Phys. B {\bf 139}, 61 (1978);
        T.N. Tovee et al., Nucl. Phys. B {\bf 33}, 493 (1971).



\bibitem{Oller01} J.A. Oller and U.-G. Mei{\ss}ner,  
        Phys.\ Lett.\  B {\bf 500}, 263 (2001).

\bibitem{Carmen02}
  C.~Garcia-Recio, J.~Nieves, E.~Ruiz Arriola, and M.~J.~Vicente Vacas,
  Phys.\ Rev.\  D {\bf 67}, 076009 (2003).

\bibitem{Jido:2003cb}
  D.~Jido, J.~A.~Oller, E.~Oset, A.~Ramos, and U.-G.~Mei{\ss}ner,
  Nucl.\ Phys.\  A {\bf 725}, 181 (2003).

\bibitem{Oller05} J.A. Oller, J. Prades, and M. Verbeni, 
        Phys.\ Rev.\ Lett.\  {\bf 95}, 172502 (2005). 

\bibitem{Borasoy:2005ie} B.~Borasoy, R.~Nissler, W.~Weise, 
        Eur.\ Phys. \ J. \ A {\bf 25}, 79 (2005).

\bibitem{Revai09}
  J. R\'evai and N.V. Shevchenko, Phys.\ Rev.\  C {\bf 79}, 035202 (2009).

\bibitem{OR98}
        E. Oset and A. Ramos, Nucl.\ Phys.\ A {\bf 635}, 99  (1998).

\bibitem{Hem} R.J. Hemingway, Nucl. Phys. B {\bf 253}, 742 (1985).

\bibitem{Tho} D.W. Thomas, A. Engler, H.E. Fisk, and R.W. Kraemer,
        Nucl. Phys. B {\bf 56}, 15 (1973).
%

\bibitem{Jido09}
        D. Jido, E. Oset, and T. Sekihara, 
	Eur. Phys. J. A {\bf 42}, 257 (2009).

\bibitem{Nacher:1998mi}
        J.C. Nacher, E. Oset, H. Toki, and A. Ramos, 
        Phys.\ Lett.\ B {\bf 455}, 55 (1999).

\bibitem{schumacher} R. Schumacher, 
        Nucl.\ Phys.\ A {\bf 835}, 231 (2010);
        K. Moriya and R. Schumacher, 
        Nucl.\ Phys.\ A {\bf 835}, 325 (2010).

\bibitem{Krehl}
	O.~Krehl, C.~Hanhart, S.~Krewald, and J.~Speth,
  	Phys.\ Rev.\ C {\bf 62}, 025207 (2000).

\bibitem{Achot}
  	A.~M.~Gasparyan, J.~Haidenbauer, C.~Hanhart, and J.~Speth,
  	Phys.\ Rev.\  C {\bf 68}, 045207 (2003).

\bibitem{Blechman}
	A.E. Blechman, A.F. Falk, D. Pirjol, and J.M. Yelton,
  	Phys.\ Rev.\ D {\bf 67}, 074033 (2003).

\bibitem{Hanhart10} C. Hanhart, Yu.S. Kalashnikova, and A.V. Nefediev, 
  Phys.\ Rev.\  D {\bf 81}, 094028 (2010).

\bibitem{Baru04}
  V.~Baru, J.~Haidenbauer, C.~Hanhart, Yu.~Kalashnikova, and A.~E.~Kudryavtsev,
  Phys.\ Lett.\  B {\bf 586}, 53 (2004).

\bibitem{Baru05}
  V.~Baru, J.~Haidenbauer, C.~Hanhart, A.~E.~Kudryavtsev, and U.-G.~Mei{\ss}ner,
  Eur.\ Phys.\ J.\  A {\bf 23}, 523 (2005).

\bibitem{Gunter01}
  J.~Gunter {\it et al.}  [E852 Collaboration],
  Phys.\ Rev.\  D {\bf 64}, 072003 (2001).

\bibitem{MHE}
  R. Machleidt, K. Holinde, and Ch. Elster, Phys. Rept. {\bf 149}, 1 (1987).

\end{thebibliography}
\end{document}